\documentclass[11pt]{article} 
\usepackage{authblk}
\usepackage[utf8]{inputenc}
\usepackage{url}
\usepackage[version=4]{mhchem}
\usepackage{multirow}
\usepackage{lipsum}
\usepackage{stackrel}
\usepackage{listings}
\usepackage{amssymb}
\usepackage{nccmath}
\usepackage{a4wide}
\usepackage{mathtools}
\usepackage{amsmath,amsfonts,amsthm,amssymb}
\usepackage{bbm, dsfont}
\usepackage[justification=justified, singlelinecheck=false]{caption}
\usepackage{graphicx}
\usepackage{enumitem}
\usepackage[justification=centering]{caption}
\usepackage{algorithmicx}
\usepackage{enumitem}
\usepackage{empheq,etoolbox}
\patchcmd{\subequations}
  {\theparentequation\alph{equation}}
  {\theparentequation.\alph{equation}}
  {}{}
\usepackage[justification=centering]{caption}
\usepackage{algorithmicx}
\usepackage{algorithm}
\usepackage{algpseudocode}
\usepackage{stackengine}

\usepackage[title]{appendix}
\usepackage[scaled=.90]{helvet}
\usepackage{courier}
\usepackage{ae}
\usepackage[T1]{fontenc}
\usepackage{graphicx}
\usepackage{subcaption}
\usepackage[export]{adjustbox}
\usepackage{wrapfig}

\usepackage[T1]{fontenc}
\usepackage{here} 
\usepackage[colorlinks=false, pdfborder={0 0 0}]{hyperref}

\usepackage{mdwlist}
\usepackage{tcolorbox}
\usepackage{fancybox}
\usepackage{framed}

\definecolor{darkblue}{rgb}{0,0,0.8}
\definecolor{darkgreen}{rgb}{0,0.8,0}

\definecolor{magenta}{rgb}{0.5,0,0.5}

\newcommand{\mathleft}{\@fleqntrue\@mathmargin0pt}

\newtheorem{definition}{Definition}[section]

\providecommand{\keywords}[1]
{
  \small	
  \textbf{\textit{Keywords---}} #1
}

\usepackage{biblatex}
\begin{document}
\sloppy
\title{American option pricing using \\ generalised stochastic hybrid systems} 
\author[1]{Evelyn Buckwar\thanks{Email: evelyn.buckwar@jku.at}}
\author[2]{Sascha Desmettre\thanks{Email: sascha.desmettre@jku.at}}
\author[3]{Agnes Mallinger\thanks{Email: agnes.mallinger@jku.at}}
\author[1]{Amira Meddah\thanks{Corresponding author: amira.meddah@jku.at}}
\affil[1]{\centerline{\small Institute of Stochastics, Johannes Kepler University Linz,} \newline \centerline{\small Altenberger Straße 69, 4040 Linz, Austria}}
\affil[2]{\centerline{\small Institute of Financial Mathematics and Applied Number Theory, Johannes Kepler University Linz,} \newline \centerline{\small Altenberger Straße 69, 4040 Linz, Austria}}
\affil[3]{\centerline{\small Linz School of Education, Johannes Kepler University Linz,} \newline \centerline{\small Altenberger Straße 69, 4040 Linz, Austria}}
\date{\today}

\newcommand{\ebbox}[1]{\fbox{$\triangleright$\textcolor{green}{\textbf{Evelyn}:} #1}}
\newcommand{\ebbbox}[1]{{
\fbox{
\parbox{0.9\textwidth}{  \fbox{$\triangleright$\asd{\textbf{Evelyn}:}} 
#1
}}}}

\newcommand{\sdbox}[1]{\fbox{$\triangleright$\textcolor{blue}{\textbf{Sascha}:} #1}}
\newcommand{\sdbbox}[1]{{
\fbox{
\parbox{0.9\textwidth}{  \fbox{$\triangleright$\asd{\textbf{Sascha}:}} 
#1
}}}}

\newcommand{\amebox}[1]{\fbox{$\triangleright$\textcolor{blue}{\textbf{Amira}:} #1}}
\newcommand{\amebbox}[1]{{
\fbox{
\parbox{0.9\textwidth}{  \fbox{$\triangleright$\asd{\textbf{Amira}:}} 
#1
}}}}

\newcommand{\agmebox}[1]{\fbox{$\triangleright$\textcolor{blue}{\textbf{Agnes}:} #1}}
\newcommand{\agmebbox}[1]{{
\fbox{
\parbox{0.9\textwidth}{  \fbox{$\triangleright$\asd{\textbf{Agnes}:}} 
#1
}}}}
\maketitle

\begin{abstract}
This paper presents a novel approach to pricing American options using piecewise diffusion Markov processes (PDifMPs), a type of generalised stochastic hybrid system that integrates continuous dynamics with discrete jump processes. Standard models often rely on constant drift and volatility assumptions, which limits their ability to accurately capture the complex and erratic nature of financial markets. By incorporating PDifMPs, our method accounts for sudden market fluctuations, providing a more realistic model of asset price dynamics. We benchmark our approach with the Longstaff-Schwartz algorithm, both in its original form and modified to include PDifMP asset price trajectories. Numerical simulations demonstrate that our PDifMP-based method not only provides a more accurate reflection of market behaviour but also offers practical advantages in terms of computational efficiency. The results suggest that PDifMPs can significantly improve the predictive accuracy of American options pricing by more closely aligning with the stochastic volatility and jumps observed in real financial markets.
\end{abstract}\vspace{0.05cm}

\keywords{American options, option pricing, Piecewise Diffusion Markov Processes, stochastic \\hybrid systems, Longstaff-Schwartz algorithm.} 

 \section{Introduction}

\indent In the complex landscape of financial markets, the accurate pricing of American options plays a pivotal role, not just for theoretical analysis but also for practical portfolio management. American options grant holders the flexibility to exercise the option prior to its maturity, a feature that significantly influences cash flows and complicates their valuation due to stochastic volatility and the need for continuous dividend considerations.\\
Traditional option pricing models, such as the Black-Scholes model, see \cite{BS73}, or the Merton model, see \cite{merton1973theory}, have been cornerstones in understanding financial derivatives. These models provide key insights for risk management and decision-making processes, allowing traders and investors to estimate the value of options under a set of simplified assumptions. However, they typically assume constant drift rates and volatility, which fails to represent the inherent unpredictability of the market and its response to external shocks such as news, economic changes, or geopolitical events.\\
Over the last two decades, it has become clear that the assumption that the underlying asset's price behaves like a geometric Brownian motion with constant drift and constant volatility cannot explain the market prices of options with different strike prices and maturities. Merton \cite{merton1976option} proposed adding jumps to the behavior of asset prices, which has led to active research into models with jumps. Several models have been proposed, including those by Kou \cite{kou2002jump} or Toivanen \cite{toivanen2008numerical}, which assume a log-double exponential distribution of jump sizes, and the Carr-Geman-Madan-Yor (CGMY) model \cite{carr2002fine}, which treats the asset price as a L\'evy process with possibly infinite jump activity. Another generalization involves stochastic volatility, as explored by Heston \cite{heston1993closed}, among others. These developments aim to capture more accurately the complex dynamics of financial markets.\\
\indent In order to incorporate these complexities into practical option pricing, several numerical methods have been developed. For instance, the binomial model has been enhanced by various researchers, including Cox et al. \cite{cox1979option}, Hull and White \cite{hull1988use}, and others, to capture the early exercise feature of options. However, these methods can be computationally intensive and memory-consuming. Similarly, Monte Carlo simulation approaches have been successful in generalizing option pricing \cite{caflisch2004monte, fu2001pricing}, although they have found limited use in scenarios involving early exercise \cite{boyle1977options, broadie1997pricing}. Furthermore, PDE methods have been notably advanced, using linear complementarity, front tracking, and front fixing methods to solve the free boundary problems associated with American options, see \cite{brennan1978finite, jaillet1990variational, arregui2020pde}.\\
A widely used method for pricing American options is the Longstaff-Schwartz (LS) algorithm \cite{LongSchw01}. This method uses Monte Carlo simulations to generate multiple potential future paths of the underlying asset's price. The value of the American option is computed by averaging the discounted cash flows from these simulated paths. Further, the optimal exercise strategy is determined by a least-squares regression approach, which estimates the continuation values (expected future payoffs if the option is not exercised) at each potential exercise date. By regressing the continuation values against a set of basis functions of the underlying asset price, the algorithm approximates the decision to hold or exercise the option.\\
\indent This paper introduces a novel approach by employing a generalised stochastic hybrid system for the pricing of American options, integrating continuous dynamics with jump processes to capture more realistic market fluctuations. Specifically, we use Piecewise Diffusion Markov Processes (PDifMPs), a type of generalised stochastic hybrid system, to model the asset price dynamics. PDifMPs combine the continuous evolution of asset prices with discrete jumps, allowing for sudden changes in market conditions, which are often triggered by unexpected events such as economic announcements or geopolitical developments.\\
\indent Over the past decades, stochastic hybrid systems (SHS) have emerged as powerful modelling techniques in various fields, including mathematical biology \cite{berg1972chemotaxis,cloez2017probabilistic}, neuroscience \cite{pakdaman2010fluid,buckwar2011exact}, biochemistry \cite{singh2010stochastic}, finance \cite{ishijima2011regime}, to name a few. A prominent class of SHS, called piecewise deterministic Markov process (PDMP), was introduced by Davis in 1984 \cite{davis1984piecewise} and involves deterministic motion punctuated by random jumps. Building on this, Blom \cite{blom1988piecewise} proposed a more general model by incorporating stochastic differential equations (SDEs) and state-dependent reset maps, leading to the formulation of PDifMPs. This framework was further developed by Bujorianu et al. in \cite{bujorianu2003reachability}, who established the theory and extended generators of PDifMPs.\\
 \indent Despite the applicability of generalised stochastic hybrid systems in various fields, including, for example, mathematical biology, \cite{meddah2023stochastic, nankep2018modelisation}, to the best of our knowledge, they have not been applied to the context of American option pricing. The important role of PDifMPs in this work lies in their ability to model the stochastic nature of financial markets, capturing both continuous trends and abrupt shifts in asset prices. In our approach, we use PDifMPs to compute the asset price at various points in time. The possible exercise times for the American options are the jump times determined by the PDifMPs. At each of these jump times, we calculate the inner value of the option. We then discount these inner values back to the present value using an appropriate discount factor. The option price is determined by taking the maximum of these inner values at the jump times, and then averaging these maximum values across all paths. Therefore, the final option price is obtained by averaging these discounted maximum values across all paths. This process is similar to the Longstaff-Schwartz algorithm in which we identify cash flows generated by the American put option at each jump time, discount these cash flows to time zero, and average them over all paths. However, instead of determining one optimal exercise time, we have the inner values at multiple stopping times.\\
The primary objective of this research is to improve the predictive accuracy of American option pricing by incorporating these advanced dynamics. By doing so, we aim to provide a model that aligns more closely with the erratic nature of financial markets, offering both theoretical insights and practical applications.\\
\indent This paper is organized as follows: Section \ref{section2} provides a concise overview of PDifMPs and the foundational tools employed in this work. In Section \ref{section3}, we introduce the mathematical framework of our proposed model, detailing the dynamic structure of asset price evolution. Section \ref{section4} explores the behavior of asset price paths through a series of numerical simulations. Section \ref{section5} focuses on American option pricing, where we present and compare results between two novel methods and the classic LS algorithm. Finally, Section \ref{sec6} concludes with a summary of our findings and suggests potential directions for future research.


\section{Fundamentals of PDifMPs}
\label{section2}
In this section, we present the fundamental concepts necessary for understanding PDifMPs. For a more complete overview, we refer the reader to \cite{bujorianu2006toward, bujorianu2003reachability, bect2007processus}.
\subsection{Definition and notations}
Consider a filtered probability space $(\Omega, \mathcal{F}, (\mathcal{F}_t)_{t \geq 0}, \mathbb{P})$, where the filtration $(\mathcal{F}_t)_{t \geq 0}$ is right-continuous and $\mathcal{F}_0$ contains all $\mathbb{P}$-null sets. Let $(W_t)_{t\in [0,T]} \in \mathbb{R}^m$, $m \in \mathbb{N}$, $T > 0$, represent a standard Wiener process defined on this space, and assume that $W_t$ is $\mathcal{F}_t$-adapted.\\
Now, consider a PDifMP $(U_t)_{t\in [0,T]} = \{ U(t,\omega) \mid t\in [0,T], \omega \in \Omega \}$, where $U_t = (S_t, \mu_t)$ takes values in $E = E_1 \times E_2$. Here, $E_1$ and $E_2$ are subsets of $\mathbb{R}^d$, $d \in \mathbb{N}$, both equipped with the Borel $\sigma$-algebra $\mathcal{B}(E)$. The set $E$ is called the state space of the process $U_t$. The PDifMP $(U_t)_{t \in [0,T]}$ consists of two components such that
\begin{itemize}
    \item The stochastic continuous component $(S_t)_{t \in [0,T]}$ has continuous paths in $E_1$.
    \item The jump component $(\mu_t)_{t \in [0,T]}$ is a process with right-continuous paths and has piecewise constant values in $E_2$.
\end{itemize}
\noindent The times at which the process $\mu_t$ undergoes a jump are denoted by $(T_i)_{i \in \mathbb{N}}$ and they form a sequence of random grid points within the interval $[0, T]$. Further, the dynamics of the PDifMP $(U_t)_{t \in [0,T]}$ on $(E, \mathcal{B}(E))$ are defined by its characteristic triple $(\phi, \lambda, \mathcal{Q})$ as follows.
\subsubsection*{Continuous dynamics}
The stochastic flow $\phi: [0,T] \times E \rightarrow E_1$, $(t,u)\mapsto \phi (t,u)$ represents the solution of a sequence of SDEs driven by $\nu$ and $\sigma$, which are the drift and diffusion coefficients respectively:
    \begin{equation}
		\label{sys_1}
		\left\{
		\begin{array}{ll}
			d\phi(t,u_i)=\nu(\phi(t,u_i),\mu_i)dt+\sigma(\phi(t,u_i),\mu_i)dW_t, \qquad t \in [T_i, T_{i+1}),\\[0.4cm]
			\phi(T_i,u_i)=s_i.
		\end{array}
		\right. 
	\end{equation}
\noindent Here, $u_i = (s_i, \mu_i)$ represents the newly updated state components at jump times $T_i$. At the endpoint $T_{i+1}$ of each interval, $s_{i+1}$ is set to the current value of $\phi(\,\cdot\,, u_i)$ to ensure the continuity of the path. Further, a new value $\mu_{i+1}$ is chosen as fixed parameter for the next interval according to the jump mechanism described below.

\subsubsection*{Jump Dynamics}
The jump dynamics are governed by the rate function $\lambda: E \rightarrow \mathbb{R}_{+}$ and transition kernel ${\mathcal{Q}:(E,\mathcal{B}(E))\rightarrow [0,1]}$, determining the frequency at which the second component of $(U_t)_{t\in [0,T]}$ jumps and the new values of the second component after a jump occurs, respectively. The process $(U_t)_{t\in [0,T]}$ experiences jumps at times $(T_i)_{i \in \mathbb{N}}$ determined by the distribution of $\lambda$ across the stochastic flow $\phi$, given by
   \begin{equation}
   \label{gen_surv}
        \mathcal{S}(t,u_i) = \exp \left(-\int_{T_i}^t \lambda(\phi(\delta, u_i), \mu_i) d\delta \right)
   \end{equation}
The function $\mathcal{S}$ is called the survival function of the inter-jump times. This function states that there is no jump in the time interval $[T_i,t)$ conditional on the process being in the initial state $u_i$.\\
The jump occurrences modify the state based on the transition probabilities specified by $\mathcal{Q}$. More precisely, let $\mathcal{U}$ be a uniformly distributed random variable on $[0,1]$, thus $\zeta:[0,1]\times E \rightarrow \mathbb{R}_{+}$ is the generalised inverse of $\mathcal{S}(t,u_i)$ defined by

\begin{equation}
   \label{genel_inv}
	\zeta(\mathcal{U},u_i)=\inf\{t\geq 0 \, ;\, \mathcal{S}(t,u_i)\leq \mathcal{U}\}.
\end{equation}

\noindent Then there exists a measurable function $\psi:[0,1]\times E \rightarrow E$ such that for $u_i\in E$ and $B\in \mathcal{B}(E)$

\begin{equation*}
	\mathbb{P}(\psi(\mathcal{U},u_i)\in B)=\mathcal{Q}(u_i,B).
\end{equation*}
The function $\psi$ represents the generalised inverse function of $\mathcal{Q}$. For a fixed $t$, $\psi(\mathcal{U}(\omega),U(\omega))$ is a random variable describing the post-jump locations of the second component of $(U_t)_{t\in [0,T]}$.

\subsection{Iterative construction of PDifMP paths}
\label{construction}
 The PDifMP $(U_t)_{t\in [0,T]}$ is constructed iteratively using its local characteristics $(\phi,\lambda, \mathcal{Q})$. Let $(\mathcal{U}_n)_{n\geq 1}$ be a sequence of iid random variables with uniform distribution on $[0,1]$ and  $u_0=(s_0,\mu_0)\in E$ the initial value of (\ref{sys_1}) at $T_0=0$, such that $u_0$ can be either an $\mathcal{F}_0$-measurable random variable (independent from the Wiener process) or a deterministic constant, for some $\omega \in \Omega$.\\
The survival function, defined in (\ref{gen_surv}), is applied, along with its generalised inverse, defined in (\ref{genel_inv}), to determine the first jump time of the second component. We proceed to define the sample path $U_t$ up to the first jump time as follows:

\begin{equation*}
	\left\{
	\begin{array}{ll}
		U_t=\phi(t, u_0) \qquad\qquad\qquad\qquad \text{for}~ 0\leq t<T_1, \\[0.2cm]
		U_{T_1}=\psi\left(\mathcal{U}_2,\big(\phi(T_1,u_0),\mu_0\big)\right).
	\end{array}
	\right. 
\end{equation*}
The trajectory of $U_t$ follows the stochastic flow $\phi$ given in (\ref{sys_1}) starting from $U_0=u_0$ until a first jump occurs at the random time $t=T_1$. The post-jump state $U_{T_1}$ is determined through the measurable function $\psi$. For all $B\in \mathcal{B}(E)$, the distribution of $\psi(\mathcal{U}_2,u_0)$ is given by

\begin{equation}
 	\mathbb{P}(\mu_{T_1}\in B\vert t=T_1, S_0=s_0)=\mathcal{Q}\left((\phi(\tau_1,u_0),\mu_0),B\right),
\end{equation}
where $\tau_1$ is the waiting time until the first jump occurs, i.e., $\tau_1=T_1$.\\
Restarting the process from the post-jump location $U_{T_1}$, we define

\begin{equation*}
	\tau_2=\zeta(\mathcal{U}_3,U_{T_1})
\end{equation*}
as the next waiting time before a jump occurs from the survival function (\ref{gen_surv}). In this way, we find the next jump time $T_2=T_1+\tau_2$.\\
Consequently, the state of the process in the interval $[T_1,T_2)$ is given by

\begin{equation*}
	\left\{
	\begin{array}{ll}
		U_t=\phi(t-T_1, U_{T_1}) \qquad\qquad\qquad\qquad\qquad \text{for}\quad T_1\leq t<T_2, \\[0.2cm]
		U_{T_2}=\psi\left(\mathcal{U}_3,\big(\phi(\tau_2,(U_{T_1},\mu_0)),\mu_0\big)\right).
	\end{array}
	\right. 
\end{equation*}
We proceed recursively to obtain a sequence of jump times $(T_i)_{i\geq 1}$, 

\begin{equation*}
	{T_i=T_{i-1}+\zeta(\mathcal{U}_{2i-1},U_{T_{i-1}})} \qquad \forall \, i\geq 1,
\end{equation*}
such that the generic sample path of $U_t$, for $t\in [T_i, T_{i+1})$, is defined accordingly by

\begin{equation*}
	\left\{
	\begin{array}{ll}
		U_t=\phi(t-T_i, U_{T_i}) \qquad\qquad\qquad\qquad\qquad\qquad \text{for}\quad T_i\leq t<T_{i+1}, \\[0.2cm]
		U_{T_{i+1}}=\psi\left(\mathcal{U}_{2i+2},\big(\phi(\tau_{i+1},(U_{T_i},\mu_{i+1})),\mu_{i+1}\big)\right).
	\end{array}
	\right. 
\end{equation*}
The number of jump times that occur between $0$ and $t$ is denoted by 

\begin{equation*}
	N_t=\sum_{i\geq 1} \mathbbm{1}_{(T_i\leq t)}.
\end{equation*}

\noindent This construction framework ensures that the sample path $U_{t}$ is defined piecewise, with each segment determined by the dynamics prescribed by $\phi$  and the transitions to new states are dictated by $\mathcal{Q}$. The following assumptions ensure that the process is a strong c\`adl\`ag \footnote{These are continuous-time stochastic processes with sample paths that are almost surely everywhere right continuous with limits from the left existing everywhere.} Markov process, guaranteeing the well-posedness of the model.

\subsection{Assumptions}
\label{assumption}
\begin{enumerate}
    \item The functions $\nu:E\rightarrow \mathbb{R}^{d_1}$ and $\sigma:E\rightarrow \mathbb{R}^{{d_1}\times m}$ are linearly bounded and globally Lipschitz continuous for all $s\in E_1$, ensuring the uniqueness solutions to the sequence of SDEs.
    \item The jump rate function $\lambda$ is measurable, such that the integral over any finite interval is finite, whereas over an unbounded interval, it is infinite, thus ensuring the existence of non-trivial dynamics.
    \item The survival function and transition kernel are measurable and probabilistically well-defined, allowing for realistic modelling of jump occurrences.
    \item For all $t>0$ and for every starting point $u_i\in E$, $\mathbb{E}[N_t\vert u=u_i]<\infty$.
\end{enumerate}
\indent Having established the fundamental principles of PDifMPs, we now apply these concepts to the modelling of American options. Our approach unfolds in three structured steps to reflect market dynamics more accurately. Initially, we model the underlying asset with a time-varying drift, improving traditional models by better capturing market fluctuations. We then elaborate on how the asset price evolves using PDifMPs, integrating both continuous movements and discrete jumps that reflect real-world financial behaviors. Then, using this modelling approach, we identify the jump times of the PDifMP that are subsequently used as potential exercise times for American options. The following section provides a detailed explanation of these steps and their implications for option pricing.
\section{Model Description: American options}
\label{section3}

\indent American options are financial derivatives that provide the holder with the right, though not the obligation, to buy or sell an underlying asset at a predetermined price (the strike price) at any time prior to the expiration date of the option. This flexibility makes the valuation of American options a more intricate process in comparison to that of their European option counterparts, which can only be exercised at maturity $T$. Here, $T$ is a positive and finite deterministic time.\\
\indent The classical approach for modelling the price of the underlying asset is based on the Black-Scholes framework \cite{BS73}. In this model, the drift rate of the underlying asset is assumed to be constant over time, thereby suggesting a fixed expected rate of return throughout the asset's lifespan. Consequently, the asset price dynamics in the classical Black-Scholes model are described by the following SDE:
\begin{equation}
	\label{Asset_price_BS}
	\left\{
	\begin{array}{ll}
		 dS_t&= \mu S_t dt + \sigma S_t dW_t \qquad \qquad t\in [0,T],\\
		S(0)&=s_0\, ,
	\end{array}
	\right. 
\end{equation}
where $\mu$ is the constant drift rate, which we usually assume to be greater than the non-negative risk-free interest rate $r$, i.e., $\mu > r$. The risk-free interest rate $r$ represents the rate of return of an investment with zero risk. In financial models, $r$ is used to discount future cash flows to their present value, reflecting the time value of money. 
\begin{definition}[Discounting factor]
The discounting factor, which is used to determine the present value of future cash flows, given a constant interest rate $r$, is defined as follows
\begin{equation}
    \label{discount_factor}
    D(t) = e^{-rt}, \qquad t\in [0,T].
\end{equation}
\end{definition}
\noindent In this context, the variable $S_t$ represents the price of the underlying asset at time $t$, $t\in [0,T]$, with initial condition $s_0$ specifying the starting price of the asset. The term $\sigma S_t$ on the r.h.s of Equation \eqref{Asset_price_BS} describes the exposure of the asset to market volatility, modeled as a function of the constant volatility parameter $\sigma$, $\sigma > 0$, associated to a standard Wiener process $W_t$.\\
The assumption of a constant drift rate in Equation \eqref{Asset_price_BS} simplifies the complex dynamics of financial markets into a more manageable form, allowing, e.g., for deriving a closed-form solution for the price of a European option. However, in reality, asset prices can experience sudden changes due to various factors such as market news, earnings announcements, or changes in economic or political conditions. These elements can induce rapid and significant fluctuations in asset prices, thereby impacting the valuation of options and various derivative instruments.\\
Acknowledging the limitations of the constant drift assumption, we expand the Black-Scholes framework to encompass assets characterised by a variable drift rate. By integrating a time-varying drift into the model, we aim to capture a more realistic behaviour of asset returns that fluctuate in response to evolving market conditions. This approach allows for a more accurate reflection of market dynamics, accommodating the unpredictability and variability inherent in financial markets.

\subsection{Modelling asset price dynamics with time-variant drift}
Having established the essential elements of our extended model, we proceed to formulate the time-varying drift SDE that captures the behaviour of the asset prices as

\begin{equation}
	\label{Asset_price}
	\left\{
	\begin{array}{ll}
		 dS_t&= \mu_t S_t dt + \sigma S_t dW_t,\\
		S(0)&=s_0.
	\end{array}
	\right. 
\end{equation}

\noindent In this formulation, the drift coefficient $\mu_t$ is subject to dynamic adjustment in order to reflect changing market conditions. It is an adapted stochastic process with respect to the filtration $(\mathcal{F}_t)_{t\in [0, T]}$, where the latter is generated by the Wiener process. We characterise $\mu_t$ as a jump process, defined as
\begin{equation}
	\label{Jump_process}
	\left\{
	\begin{array}{ll}
		 d\mu_t &= 0 \, dt, \\
		\mu(0) &= \mu_0.
	\end{array}
	\right. 
\end{equation}
Here, $\mu_t$ takes piecewise constant values, changing only at discrete jump times $T_i$, $T_i<T$. More precisely, on each interval $[T_i, T_{i+1})$, $i = 1, \dots, n$, $\mu_t$ remains constant but may take different values across intervals, reflecting the occurrence of jumps. The mechanism by which the jumps occur and thus $\mu_t$ takes new values will be specified in the following section.\\
To model the asset price evolution under the influence of the dynamic drift, we assume that the sample path of an asset starting with value $s_0$ evolves according to SDE \eqref{Asset_price}, influenced by the dynamic drift defined in Equation \eqref{Jump_process} for random periods. To describe this process, we rely on the PDifMP framework introduced in Section \ref{section2}. We set $E_1 = \mathbb{R}$, $E_2=\mathbb{R}_{+}$ and denote by $U_t := (S_t, \mu_t)$ the PDifMP describing asset price dynamics. Here, the continuous process $S_t$ and hence the process $U_t$ are affected by spontaneous drift changes induced by the jump process $\mu_t$. Thus, the state space of the piecewise process $U_t = (S_t, \mu_t)$ for asset price dynamics is $E = E_1 \times E_2$. The solution to the coupled system \eqref{Asset_price}-\eqref{Jump_process} is denoted by $\phi(t, U_t):=\phi (t, (S_t,\mu_t))$.\\
\noindent Throughout this paper, "jumps" represent sudden and significant changes in the price of the underlying asset, which can be either positive or negative. These jumps capture real-world scenarios where asset prices experience abrupt shifts due to market events. Moreover, the "jump times", correspond to the moments when these significant price changes occur and represent potential exercise dates for the American option. These times are particularly relevant when the price of the underlying asset is favourable - for example, when it exceeds the strike price of a call option - and thus influences the optimal exercise strategy. 

\subsection{Characterisation of the drift coefficient as a jump process} 
\label{Charct_of_the_drift}
\indent In this section, we focus on the PDifMP $U_t = (S_t, \mu_t)$ and define the jump mechanism of this process. Specifically, we detail the last two components of the characteristic triplet of the PDifMP $U_t$,  namely the jump rate function $\lambda$ and the transition kernel $\mathcal{Q}$, which together characterise the jump mechanism of the process. Indeed, the definition of the jump mechanism is crucial for capturing the exercise decisions of American options. These options introduce significant challenges in valuation and management due to the flexibility of early exercise, particularly when considering the optimal exercise timing based on evolving market conditions and the price performance of the underlying.\\
In our model, the jumps of the PDifMP $U_t$ are governed by a non-homogeneous Poisson process with an intensity function $\lambda(U_t) := \lambda(S_t, \mu_t)$. This function allows the jump frequency to dynamically depend on both the current asset price and drift, reflecting the complex interplay between market conditions and strategic option holder decisions. More precisely, we choose
\begin{equation}
    \lambda(U_t) := \lambda_0 + \eta \max\left(0, (|S_t - \delta| - \beta)\right), \qquad t\in [0,T].
    \label{jump_Rate}
\end{equation}
\noindent Here, $\delta$ is a reference point for the asset price $S_t$ around which the sensitivity of the jump rate $\lambda(U_t)$ is evaluated. Depending on the context, $\delta$ can be set to the strike price $K$, the initial asset price $S_0$, or some other reference value. For instance, when $\delta$ is set to $K$, the model focuses on the deviation of the asset price from the strike price, which is important in assessing the moneyness of the option. When $\delta$ is set to $S_0$, the model captures the volatility relative to the initial price of the asset. Therefore, $\delta$ could represent any specific threshold or benchmark around which the sensitivity of the jump rate is evaluated.\\
The parameter $\lambda_0$ represents the minimum intensity or frequency at which jumps in $\mu_t$ will occur under normal conditions. Essentially, $\lambda_0$ reflects the inherent volatility or instability of the market conditions surrounding the asset, independent of extreme price movements. This could be seen as capturing the background noise of the market, i.e. frequent but smaller shifts in market dynamics that are always present, even when no significant market events are driving significant changes.\\
\noindent Further, the role of the parameters $\beta$ and $\eta$ is to control how sensitive the jump rate is to changes in the asset price relative to the reference point $\delta$. More specifically, $\beta$ acts as a buffer zone around $\delta$. This means that within the buffer zone $(|S_t - \delta| \leq \beta)$, the jump rate remains at its baseline level $\lambda_0$, reflecting normal market activity where small fluctuations in the asset price near $\delta$ are unlikely to trigger significant changes. However, outside the buffer zone $(|S_t - \delta| > \beta)$, the jump rate begins to increase with distance from $\delta$, reflecting increased market sensitivity and activity. As the asset price moves further away from $\delta$, the probability of significant market events increases, potentially influencing decisions such as the exercise of options. This increase in jump rate is scaled by the parameter $\eta$.\\
\noindent In this context, $\eta$ acts as a scaling factor that adjusts the jump intensity function $\lambda(U_t)$, directly affecting how the jump rate responds to deviations in the asset price $S_t$ from $\delta$. In particular, once $S_t$ exceeds the buffer $\beta$, each unit of price deviation increases the jump rate by $\eta$. Higher $\eta$ values increase the sensitivity of the jump rate, making the model highly responsive in volatile markets where stock prices are prone to dramatic shifts. This responsiveness allows the model to adapt rapidly to changes, reflecting the rapid actions of market participants. Conversely, a lower $\eta$ reduces the sensitivity of the model, stabilising the jump rate in less volatile markets where price changes are gradual and predictable, preventing overreactions to minor fluctuations and better suited to environments with less reactive market behaviour.\\
\indent We now introduce the transition kernel $\mathcal{Q}$ of the process $U_t$ which, together with the jump rate $\lambda$, define the jump mechanism of the process. In particular, we employ the Laplace distribution as the probability density function for the drift dependent on the current state of the asset $S_t$, given by
\begin{equation}\label{Lap_dis}
    \mathcal{Q}(U_t):= f(\mu | a(S_t), b) = \frac{1}{2b} \exp\left(-\frac{| \mu - a(S_t)|}{b}\right),
\end{equation}

where
\begin{equation}
   a(S_t) = \mu_0 + \alpha(S_t - \delta) \,.
   \label{a_EQ}
\end{equation}

\noindent Equation \eqref{a_EQ} acts as the location parameter dependent on the price of the asset. The parameter $\mu_0$ serves as a baseline for the drift level, and $\alpha(S_t - \delta)$ adjusts this base level based on the deviation of asset price from $\delta$. Moreover, $\alpha$ in Equation \eqref{a_EQ} is a scaling factor that determines how the expected drift level, $\mu_t$, adjusts in response to changes in the asset price relative to the reference point $\delta$. It quantifies the extent to which the drift is influenced by the asset price moving away from or towards $\delta$. In practical terms, the parameter $\alpha$ helps financial analysts and traders predict how sensitive the drift (and thus the asset or option pricing) is to changes in the market environment, aiding in more informed strategic decisions.\\
\noindent Further, the parameter $b$ in \eqref{Lap_dis} denotes the scale parameter of the Laplace distribution capturing the dispersion around the mean location $a(S_t)$. In particular, the choice of $b$ affects the sensitivity of the model to underlying market dynamics. For instance, in volatile markets, a larger $b$ value might be appropriate to capture the wide variance in drift changes, whereas in more stable markets, a smaller 
value for $b$ could suffice. Hence, when pricing options, the choice of $b$ is important as it affects the valuation of risk and uncertainty in future returns, directly influencing pricing strategies and risk assessments.\\
\noindent The Laplace distribution, known for its peakedness and heavy tails compared to the normal distribution, is particularly adept at representing significant yet infrequent events that can heavily impact financial markets. Therefore, the choice of a Laplace distribution to model the jumps associated with the drift process $\mu_t$ captures the asymmetric nature of financial market responses to various types of news and the leverage effect, where volatility increases as prices decrease. The ability of this distribution to represent skewed outcomes facilitates a more realistic representation of market responses to shocks, allowing for potentially larger changes in $\mu_t$ depending on the relationship between $S_t$ and $\delta$. Furthermore, the heavy tails of the Laplace distribution accommodate the fat-tailed behaviour of financial returns, acknowledging the higher likelihood of extreme market movements.\\
\indent Finally, using the definition of the survival function (\ref{gen_surv}), it is possible to construct the sequence of jump times $(T_n)_{n\geq 1}$, with ${T_n=\tau_1+\dots+\tau_n}$, for all $n\geq 1$ (and $T_0=0$ by convention), such that the process $U_t$ describing the asset price characterised by a variable drift rate, is piecewise constructed on each interval $[T_i,T_{i+1})$, $i=1,\dots,n$, via the characteristics $(\phi,\lambda,\mathcal{Q})$ given by

\begin{equation}
	\label{charc}
	\left\{
	\begin{array}{ll}
		\phi& = S_0\exp\left( (\mu_t- \frac{\sigma^2}{2})t + \sigma W_t\right),\\[0.2cm]
		\lambda & = \lambda_0 + \eta \max\left(0, (|S_t - \delta| - \beta)\right),\\[0.2cm]
		\mathcal{Q} & = \frac{1}{2b} \exp\left(-\frac{| \mu - a(S_t)|}{b}\right) \,.
	\end{array}
	\right. 
\end{equation}
Here, $\phi$ is the solution of the coupled system \eqref{Asset_price}-\eqref{Jump_process} and $\mu_t$ is a piecewise constant over each interval of random length $T_{i+1}-T_i$. As proven in \cite{bujorianu2006toward}, under the assumptions given in Section \ref{assumption}, this construction leads to a c\`adl\`ag strong Markov process, describing in our context the behaviour a stock price under market fluctuations.\\
\noindent Therefore, the system describing the behaviour of a stock price under market fluctuations is thus a system concatenated over all subsequent intervals $[T_i,T_{i+1})$, such that for all ${t\in [T_i,T_{i+1})}$, $i\geq 0$, we have

\begin{equation}
	\label{Micro_sys}
	\left\{
	\begin{array}{ll}
		dS_t & = \mu_t S_t dt + \sigma S_t dW_t,\\[0.2cm]
		d\mu_t & = 0dt,\\
	\end{array}
	\right. 
\end{equation}

\noindent with initial values at the jump time $T_i$ given by the values of $S_t$ at the endpoint of the previous interval and a new constant value for $\mu_t$ drawn from the transition kernel $\mathcal{Q}$.
The overall process of concatenated solutions of \eqref{Micro_sys} is a couple $U_t=(S_t,\mu_t) \in E$, with ${E=\mathbb{R}\times \mathbb{R}_{+}}$. 

\section{Investigation of the asset price paths}
\label{section4}

Understanding the evolution of asset prices over time is crucial for accurately pricing options and managing financial risk. In this section, we explore the dynamics of asset price movements using a PDifMP model that incorporates jump processes to capture sudden market shifts. The PDifMP model extends traditional approaches by allowing for time-varying drift and jump intensity, providing a more flexible and realistic framework for modelling asset prices.\\
The importance of this research lies in its potential to improve our understanding of how different market conditions and model parameters affect the behaviour of asset prices. By examining the interplay between the baseline jump intensity $ \lambda_0 $, the sensitivity parameter $ \eta $ and the initial asset price $ S_0 $, we aim to gain insights into how these factors influence price paths, volatility and the frequency of significant market movements.\\
This section is structured around a series of numerical experiments designed to explore the effects of these parameters on asset price evolution. These experiments are essential to validate the PDifMP model's ability to simulate realistic asset price behaviour, particularly in comparison with traditional models such as Black-Scholes, which assume constant drift. 
\subsection{Numerical simulations}
Let $T$ be a fixed time horizon. In this section, we conduct a series of simulations of the PDifMP $U_t$, which models the evolution of the asset price as defined in \eqref{Micro_sys}. The purpose of these simulations is to examine the impact of the jump rate function on significant changes (both positive and negative) in the asset price. The simulation of the PDifMP $U_t$ is based on the construction detailed in Section \ref{construction} and is carried out using a self-developed code in $\mathbf{R}$ (See Algorithm \ref{alg:assetPricePDifMP}). For more details about the simulation of PDifMPs, we refer the reader to \cite{buckwar2024numerical}.

\begin{algorithm}[H]
\caption{Asset price as PDifMP}
\begin{algorithmic}[1]
\Require Intensity $\lambda$, maturity $T$, initial asset price $S_0$.
\State Initialise vector $S$ with $S_0$.
\State Initialise vector $jumpTimes$ with 0.
\State Generate $T_1$ by the thinning method and rate $\lambda$ and $S_{new}$ until $T_1$ as a solution of the corresponding SDE.
\State Set a counter $i = 1$.
\While{$T_i < T$}
    \State Append $S_{new}$ to $S$.
    \State Append $T_i$ to $jumpTimes$.
    \State Draw a new $\mu$ from the Laplace distribution.
    \State $i = i + 1$.
    \State Generate $T_i$ by the thinning method with rate $\lambda$ and
        simulate $S_{new}$ until $T_i$ as a solution of the corresponding SDE.  
\EndWhile
\State Simulate $S_{new}$ until $T$ as a solution of the corresponding SDE.
\State Append $S_{new}$ to $S$.
\State Append $T$ to $jumpTimes$.
\State Output $S$ and jump times $jumpTimes$.
\end{algorithmic}
\label{alg:assetPricePDifMP}
\end{algorithm}
\noindent For this purpose, we first specify in Table \ref{parameter_mod1} the parameters and coefficient involved in the simulation of the System \eqref{Micro_sys}.
\begin{table}[h]
\begin{center}
\begin{tabular}{c|c|c} 
    \hline  
    \rule{0pt}{3ex}Parameter & Description & Value (unit) \\[1ex]
    \hline
    \rule{0pt}{2ex} $s_0$ & initial stock price    & $36\,\$$ \\[1.5ex]
    \rule{0pt}{2ex} $\lambda_0$  & minimum frequency of fluctuation &$0.01-50$ \\[1.5ex] 
    \rule{0pt}{2ex} $\eta$  & sensitivity of fluctuation in $\lambda(U_t)$ &$0-1$\\[1.5ex]
    \rule{0pt}{2ex} $K$ & strike price & $40\,\$$\\[1.5ex]
    \rule{0pt}{2ex} $\mu_0$ & initial drift  & $0.06$   \\[1.5ex]
    \rule{0pt}{2ex} $\sigma$ & diffusion coefficient  & $0.2$   \\[1.5ex]
    \rule{0pt}{2ex} $\alpha$ & scaling factor & $10^{-6}$   \\[1.5ex]
    \rule{0pt}{2ex} $b$ & scaling factor & $0.01$   \\[1.5ex]
    \rule{0pt}{2ex} $\beta$ & buffer zone value & $0$   \\[1.5ex]
    \hline
\end{tabular}
\end{center}
\vspace{0.3cm}
\caption{\textbf{Model parameters.} Parameters used for simulating PDifMP asset prices.}
\label{parameter_mod1}
\end{table}
\noindent We now present a series of numerical experiments to obtain insight into several features characterising the proposed approach. More precisely,
\begin{itemize}

  \item[{\bf(A)}]  we consider the model for fixed values for $\lambda_0$, $\alpha$ and $b$ and we assess the effects of varying $\eta$ on the stock price;
  \item[{\bf(B)}] we consider the model for fixed values for $\eta$, $\alpha$ and $b$, and we evaluate the effects of the variation of $\lambda_0$ on the stock price evolution and the potential exercise times.
    \item[{\bf(C)}] we fix the values of $\alpha$ and $b$ and consider different combinations of $\lambda_0$ and $\eta$ to show how their respective effects merge.
\end{itemize}
\indent Starting with the numerical experiment {\bf(A)}, we examine the effects of varying $\eta$. This experiment is motivated by the need to understand the sensitivity of the stock price to changes in $\eta$, as it directly adjusts the responsiveness of the model to price deviations from the reference point $\delta$, thereby influencing the overall dynamics of asset prices. We note that precise estimation of $\lambda_0$, $\alpha$ and $b$ can be difficult.  
As detailed in Section \ref{Charct_of_the_drift}, $\lambda_0$ represents the baseline jump intensity, while $\alpha$ and $b$ are scaling parameters in the Laplace distribution. 
\\
We fix $\lambda_0 = 5$; the remaining parameters are chosen as in Table \ref{parameter_mod1}, and we investigate the asset price evolution for different values of $\eta \in [0,1]$.
\begin{figure}[H]
    \centering
   \begin{subfigure}[t]{0.45\textwidth}
        \centering
        \includegraphics[width=\textwidth]{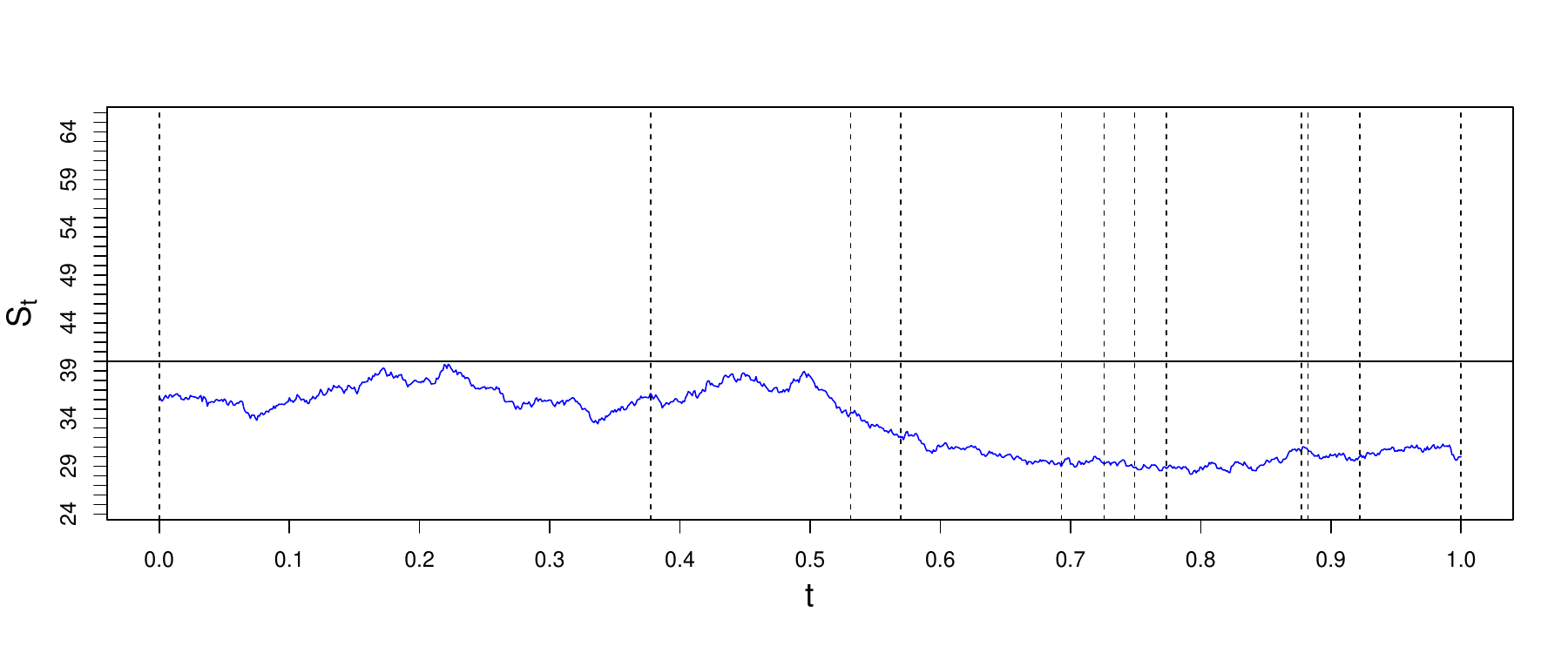}
        \caption{Asset price evolution for $\eta = 0$.}
    \end{subfigure}
    \hfill
    \begin{subfigure}[t]{0.45\textwidth}
        \centering
        \includegraphics[width=\textwidth]{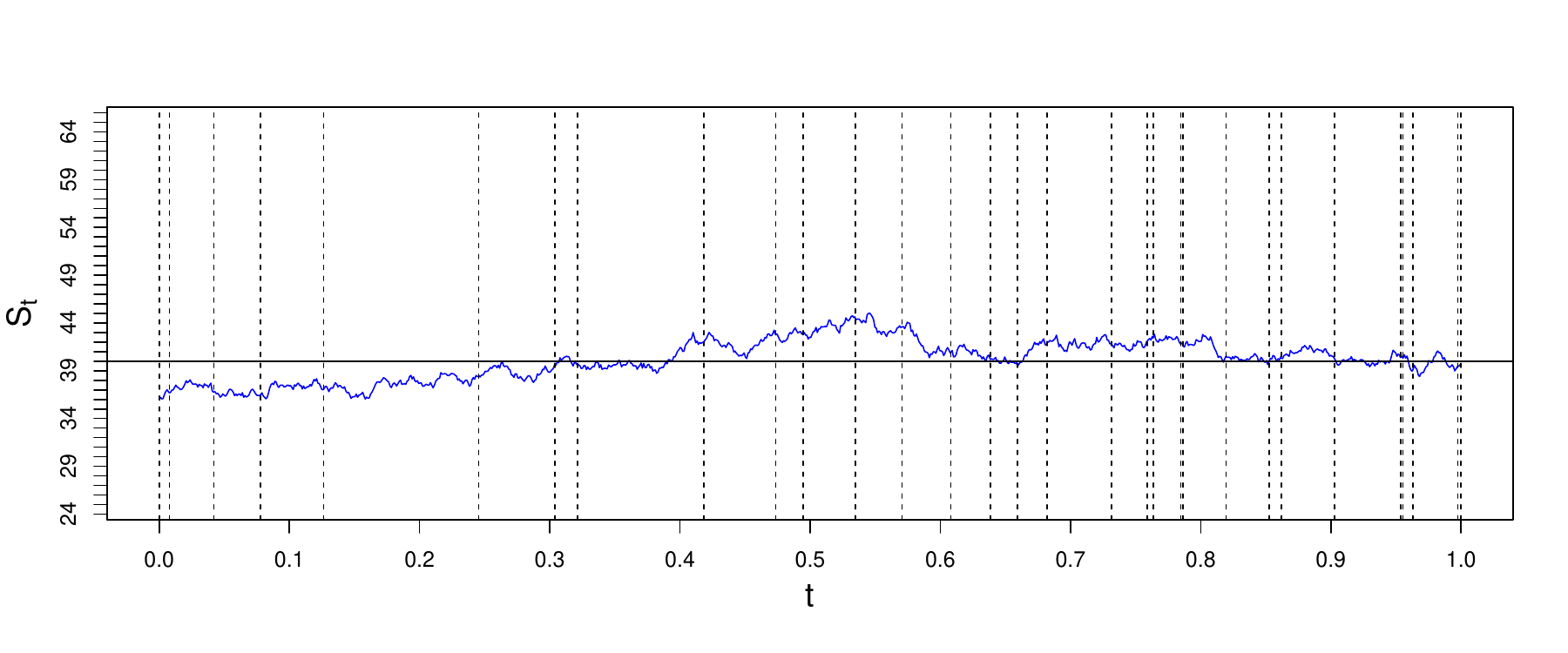}
        \caption{Asset price evolution for $\eta = 0.3$.}
    \end{subfigure}

    \vspace{0.5cm}

    \begin{subfigure}[t]{0.45\textwidth}
        \centering
        \includegraphics[width=\textwidth]{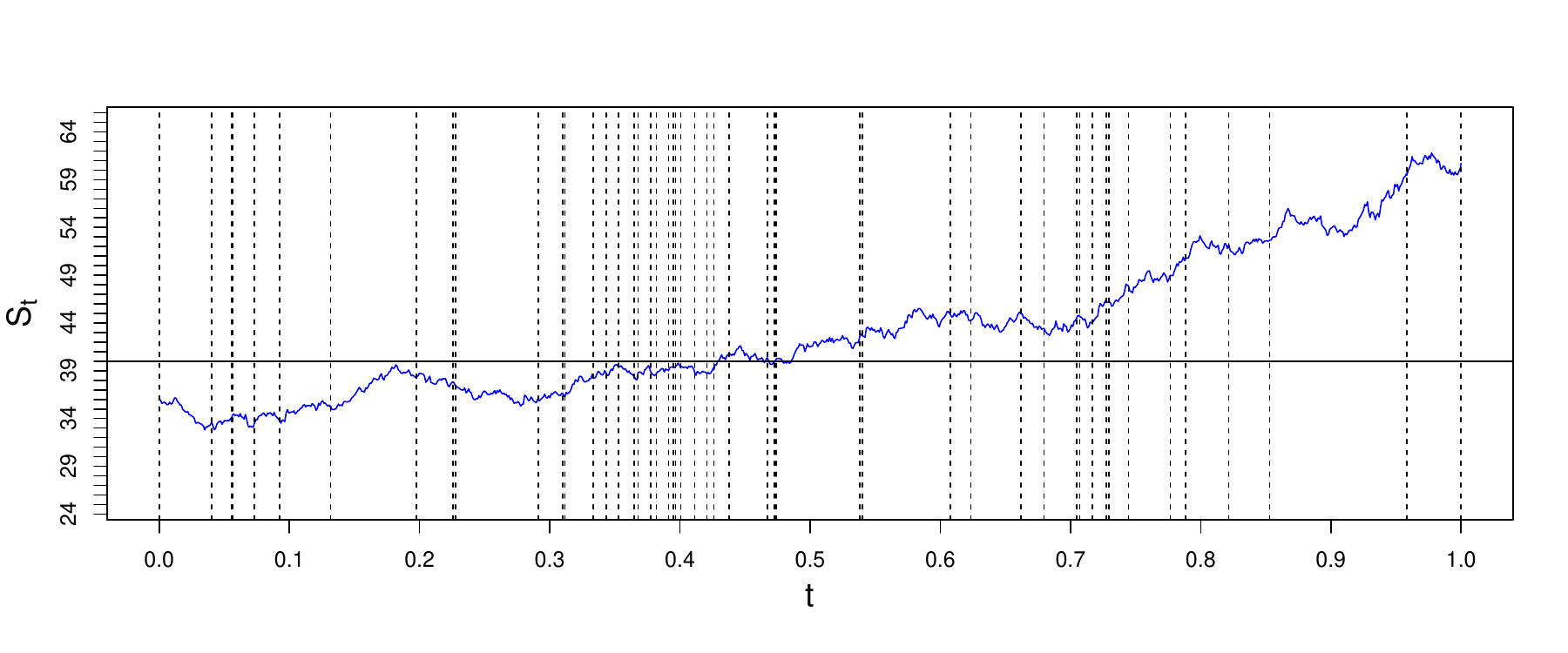}
        \caption{Asset price evolution for $\eta = 0.5$.}
    \end{subfigure}
    \hfill
    \begin{subfigure}[t]{0.45\textwidth}
        \centering
        \includegraphics[width=\textwidth]{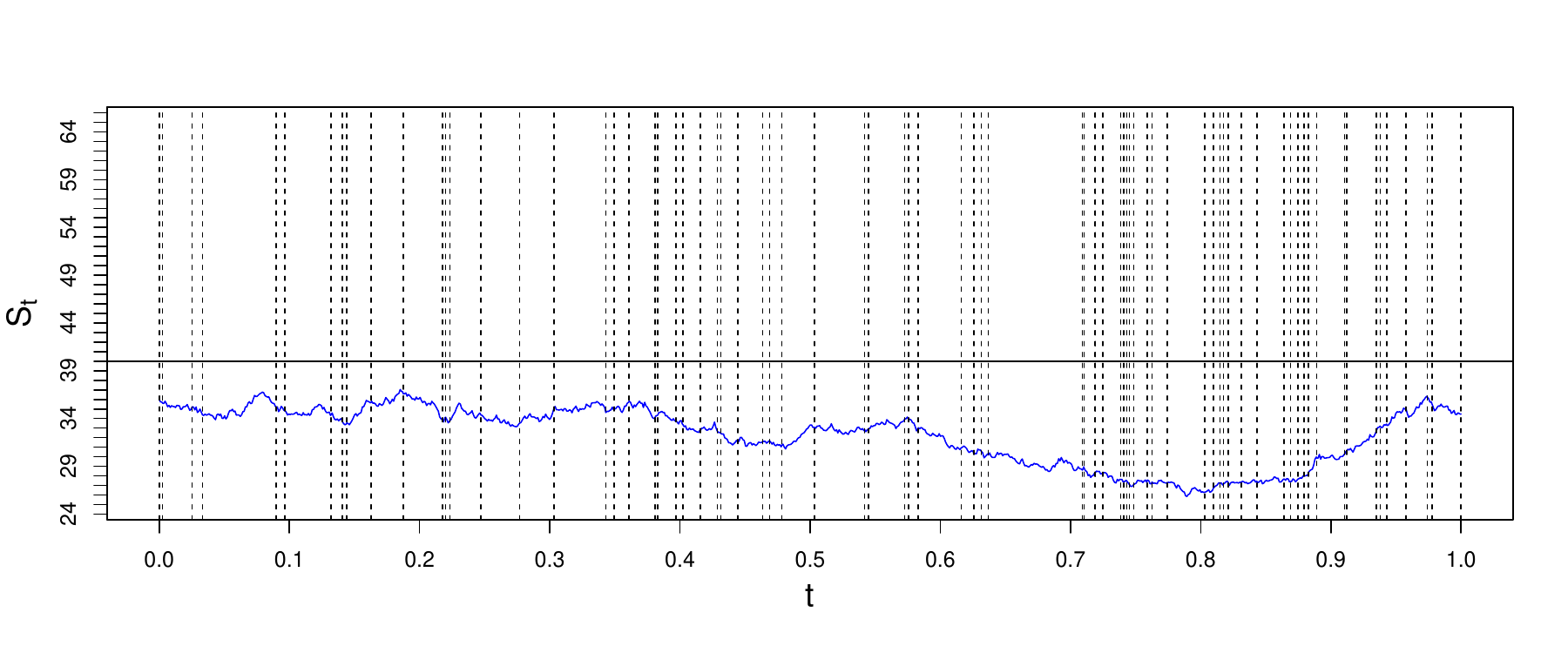}
        \caption{Asset price evolution for $\eta = 1$.}
    \end{subfigure}

    \caption{{\bf Scenario (A).} Numerical simulations of Equation \eqref{Micro_sys} with the parameters listed in Table \ref{parameter_mod1}, $\lambda_0=5$ and for different values of $\eta$. The asset price evolution is shown after 1 year. The horizontal line represents the strike price.}
\label{scenario_A}
\end{figure}

\begin{figure}[H]
    \centering

    \begin{subfigure}[t]{0.45\textwidth}
        \centering
        \includegraphics[width=\textwidth]{lambda5_eta1.pdf}
        \caption{Asset price evolution for $\lambda_0 = 5$.}
    \end{subfigure}
    \hfill
    \begin{subfigure}[t]{0.45\textwidth}
        \centering
        \includegraphics[width=\textwidth]{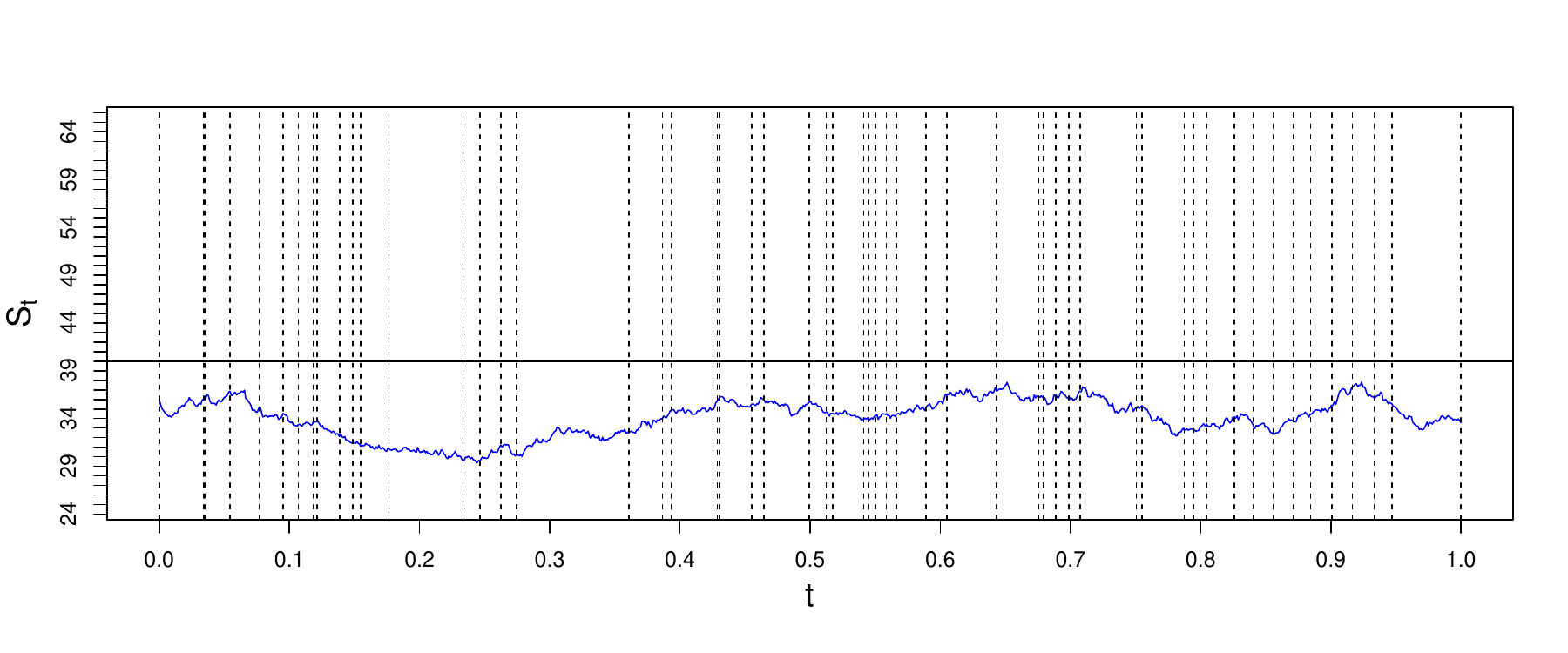}
        \caption{Asset price evolution for $\lambda_0 = 1$.}
    \end{subfigure}

    \vspace{0.5cm}

    \begin{subfigure}[t]{0.45\textwidth}
        \centering
        \includegraphics[width=\textwidth]{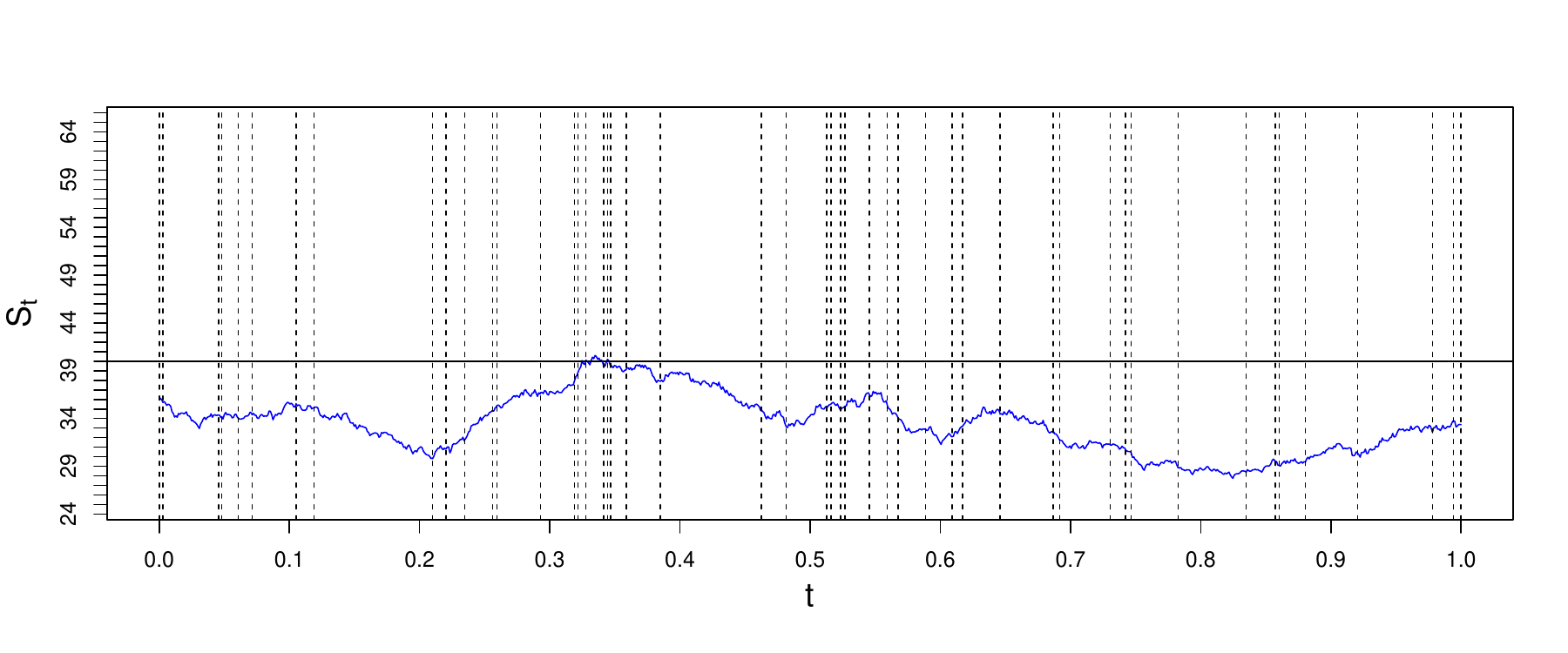}
        \caption{Asset price evolution for $\lambda_0 = 0.1$.}
    \end{subfigure}
    \hfill
    \begin{subfigure}[t]{0.45\textwidth}
        \centering
        \includegraphics[width=\textwidth]{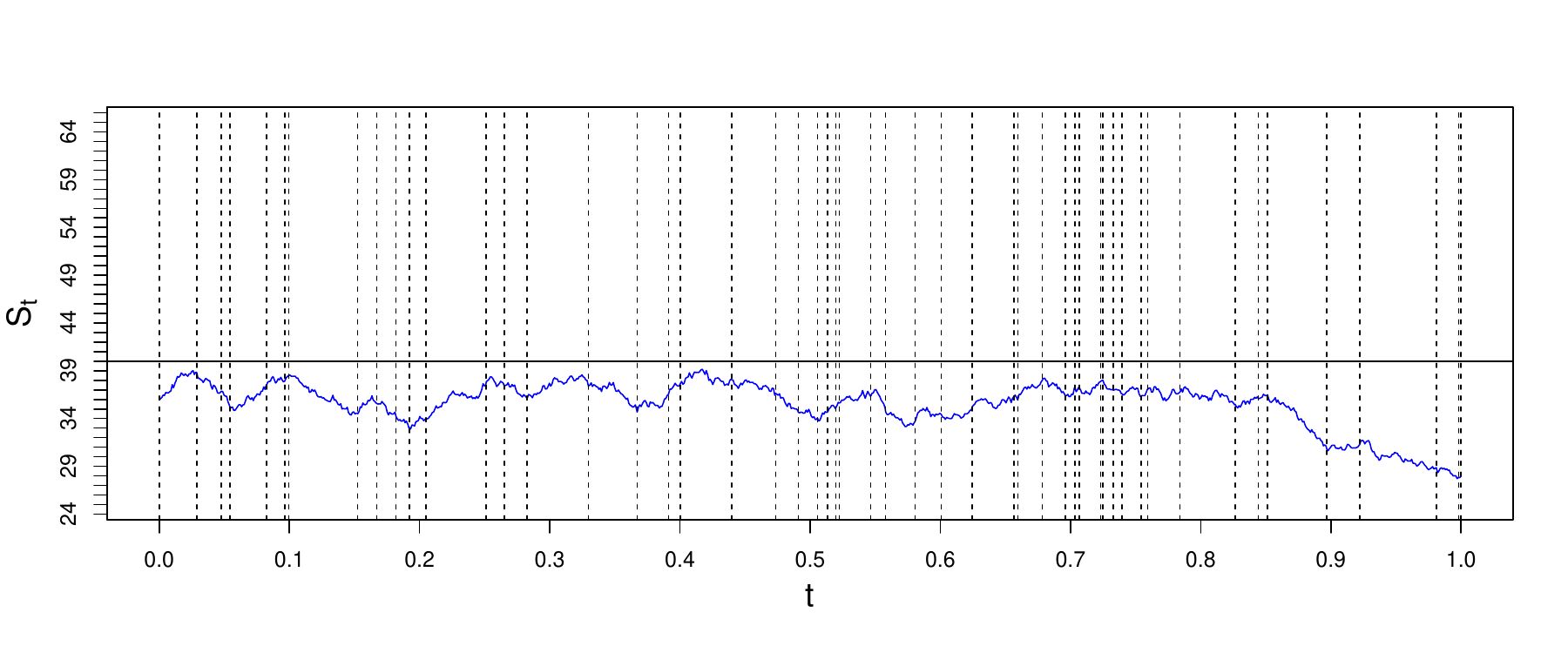}
        \caption{Asset price evolution for $\lambda_0 = 0.01$.}
    \end{subfigure}

    \caption{{\bf Scenario (B).} Numerical simulations of Equation \eqref{Micro_sys} with the parameters listed in Table \ref{parameter_mod1}, here $\eta=1$ is fixed, and we test for different values of $\eta$. The asset price evolution is shown after 1 year. The horizontal line represents the strike price.}
\label{scenario_B}
\end{figure}

\noindent In Figure \ref{scenario_A}, the plots illustrate the evolution of the asset price over a one-year period under different market sensitivities to price deviations from the initial asset price $S_0$. In subplot (a), where $\eta = 0$, the asset price follows a relatively smooth trajectory with 12 jumps, indicating minimal sensitivity to deviations from $S_0$. As $\eta$ increases to 0.3 in subplot (b), the number of jumps increases to 32, leading to slightly more volatility, although the overall price path remains stable. Subplot (c) with $\eta = 0.5$ shows a further increase in sensitivity, resulting in 52 jumps and a more pronounced upward trend in the asset price. Finally, in subplot (d) with $\eta = 1$, the asset price shows 78 jumps, but the trend is less clear, with the path showing more frequent but not necessarily larger deviations from $S_0$. This suggests that while $\eta$ controls the frequency of jumps in the process, it does not directly affect the smoothness or volatility of the asset price path.\\
\noindent For the numerical test {\bf(B)}, we fix $\eta = 1$ and vary the value of the parameter $\lambda_0$, which relates to the inherent volatility of the market conditions surrounding the asset. The results of the simulations for ${\lambda_0 \in [0.01, 5]}$ are shown in Figure \ref{scenario_B}.

\begin{figure}[H]
    \centering

    \begin{subfigure}[t]{0.45\textwidth}
        \centering
        \includegraphics[width=\textwidth]{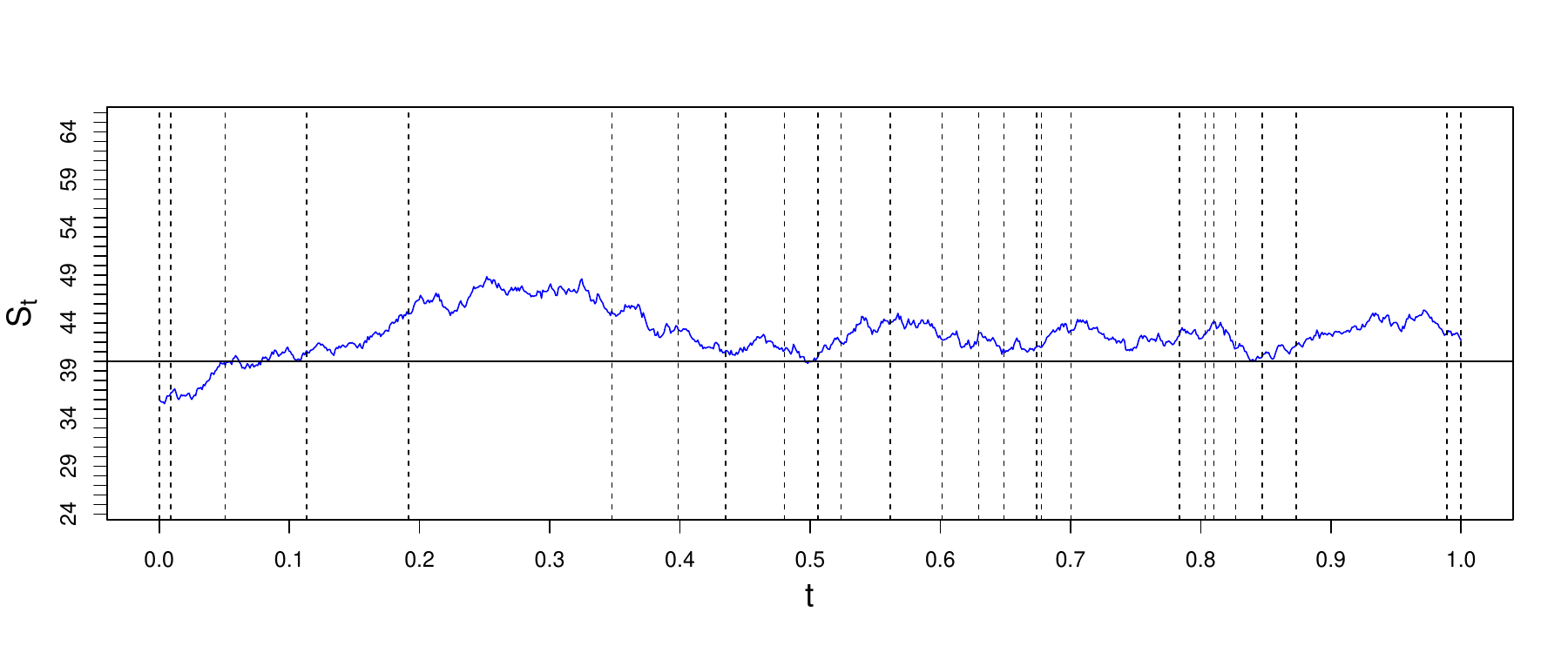}
        \caption{Asset price evolution for $\lambda_0 = 20$ and $\eta=0$.}
    \end{subfigure}
    \hfill
    \begin{subfigure}[t]{0.45\textwidth}
        \centering
        \includegraphics[width=\textwidth]{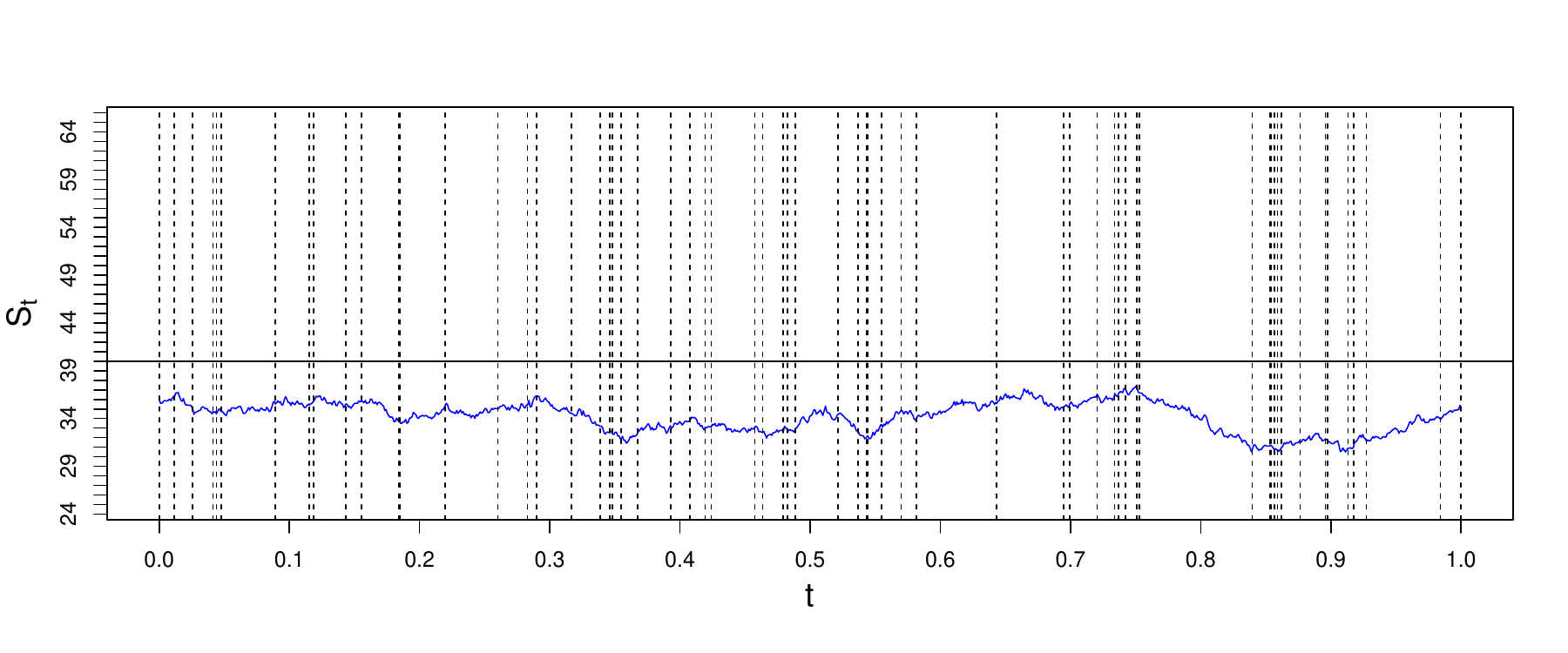}
        \caption{Asset price evolution for $\lambda_0 = 50$ and $\eta=0.3$.}
    \end{subfigure}

    \caption{{\bf Scenario (C).} Numerical simulations of Equation \eqref{Micro_sys} with the parameters listed in Table \ref{parameter_mod1}. Here, we consider different values of $\eta$ and $\lambda_0$. The asset price evolution is shown after 1 year. The horizontal line represents the strike price.}
\label{scenario_C}
\end{figure}

\noindent The plots in Figure \ref{scenario_B} show the asset price evolution over one year under varying baseline jump intensities. In subplot (a), where $\lambda_0 = 5$, the asset price exhibits significant volatility with 78 jumps. Subplot (b) with $\lambda_0 = 1$ shows a similar volatility pattern. As $\lambda_0$ decreases to $0.1$ in subplot (c) and $0.01$ in subplot (d), the asset price trajectories still display high volatility and 52 jumps. This indicates that for a high sensitivity parameter $\eta = 1$, changing the baseline jump intensity $\lambda_0$ does not significantly influence the volatility and frequency of jumps in the asset price. This occurs because the high value of $\eta$ makes the jump intensity highly responsive to deviations in the asset price, overshadowing the effects of changes in $\lambda_0$. Consequently, even with a lower $\lambda_0$, the high $\eta$ maintains substantial volatility due to its strong reaction to price deviations.\\
\noindent Referring to test {\bf (C)}, we analyse the interplay between the effects of the parameters $\lambda_0$ and $\eta$. In particular, we consider two different combinations $\eta=0$, $\lambda_0=20$ and $\eta=0.3$, $\lambda_0=50$. The results are shown in Figure \ref{scenario_C}.

\noindent In Figure \ref{scenario_C}, we explore the asset price evolution over one year under two different scenarios: (a) with $\lambda_0 = 20$ and $\eta = 0$, and (b) with $\lambda_0 = 50$ and $\eta = 0.3$. In subplot (a), where the initial asset price is set to $36\,\$$, the asset price shows a clear upward deviation from the initial value, reflecting a relatively smooth upward trend. In subplot (b), despite the higher $\lambda_0$ of 50 and a moderate $\eta = 0.3$, the asset price remains relatively stable and close to the initial value, without a significant upward or downward trend. Interestingly, while subplot (a) exhibits a positive trend, subplot (b) does not show increased volatility in terms of the asset price fluctuations compared to subplot (a). Instead, the main difference between the two scenarios lies in the number of jumps: subplot (b) has a higher frequency of jumps due to the increased $\lambda_0$, which indicates more frequent but smaller adjustments to the asset price. This suggests that the increased jump rate in subplot (b) introduces more frequent corrections, keeping the asset price close to its initial value, while in subplot (a), the absence of jump sensitivity allows for a more pronounced deviation from the starting price.\\
\noindent We now recall the jump rate function as detailed in \eqref{jump_Rate}, and we set $\delta = S_0$, the initial asset price, which allows us to examine the behaviour of the asset price under different initial conditions relative to the strike price.\\ 
To structure our analysis, we consider two primary scenarios based on initial values of $S_0$: one where $S_0 = 36$, which is below the strike price, and another where $S_0 = 44\,\$$, which is above the strike price. For each of these scenarios, we conduct a series of numerical simulations to assess how the asset price evolves under different combinations of $\lambda_0$ and $\eta$. Hereafter, it is important to note that all the asset price paths in the following scenarios are generated using the same Wiener process, ensuring that variations in the paths are solely due to the differences in the parameters $\eta$ and $\lambda_0$. These scenarios are outlined as follows:
\begin{itemize}
    \item[\bf(D)] We examine the evolution of the asset price for an initial stock price $S_0 = 36\,\$$, under various values of $\lambda_0$ and $\eta$ to observe how the model responds when the initial price is below the strike price.
    \item[\bf(E)] We repeat the analysis with an initial stock price $S_0 = 44\,\$$, testing the same range of $\lambda_0$ and $\eta$ values to compare how the price dynamics differ when the initial price is above the strike price.
\end{itemize}

\begin{figure}[H]
    \centering

    \begin{subfigure}[t]{0.45\textwidth}
        \centering
        \includegraphics[width=\textwidth]{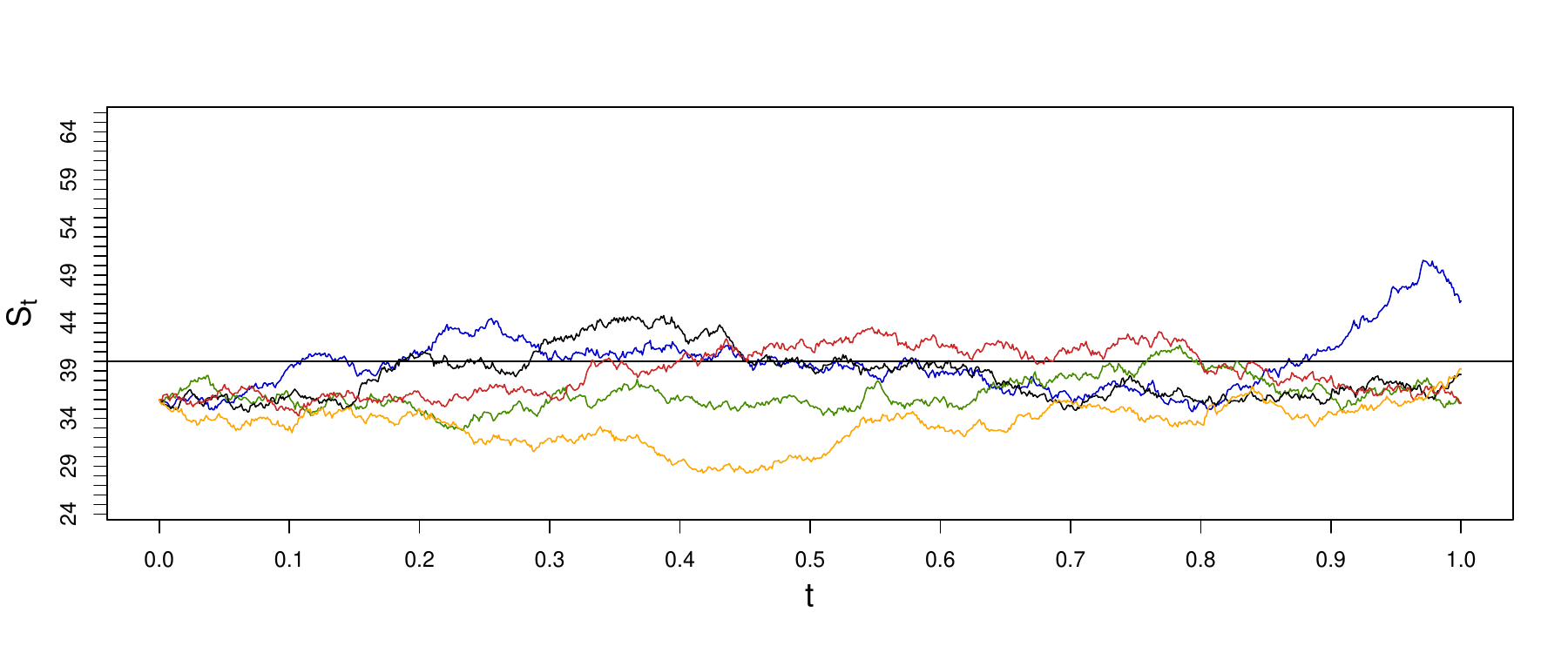}
        \caption{Asset price evolution for initial $S_0=36$ and $\lambda_0 = 10$ and $\eta=0$.}
    \end{subfigure}
    \hfill
     \begin{subfigure}[t]{0.45\textwidth}
        \centering
        \includegraphics[width=\textwidth]{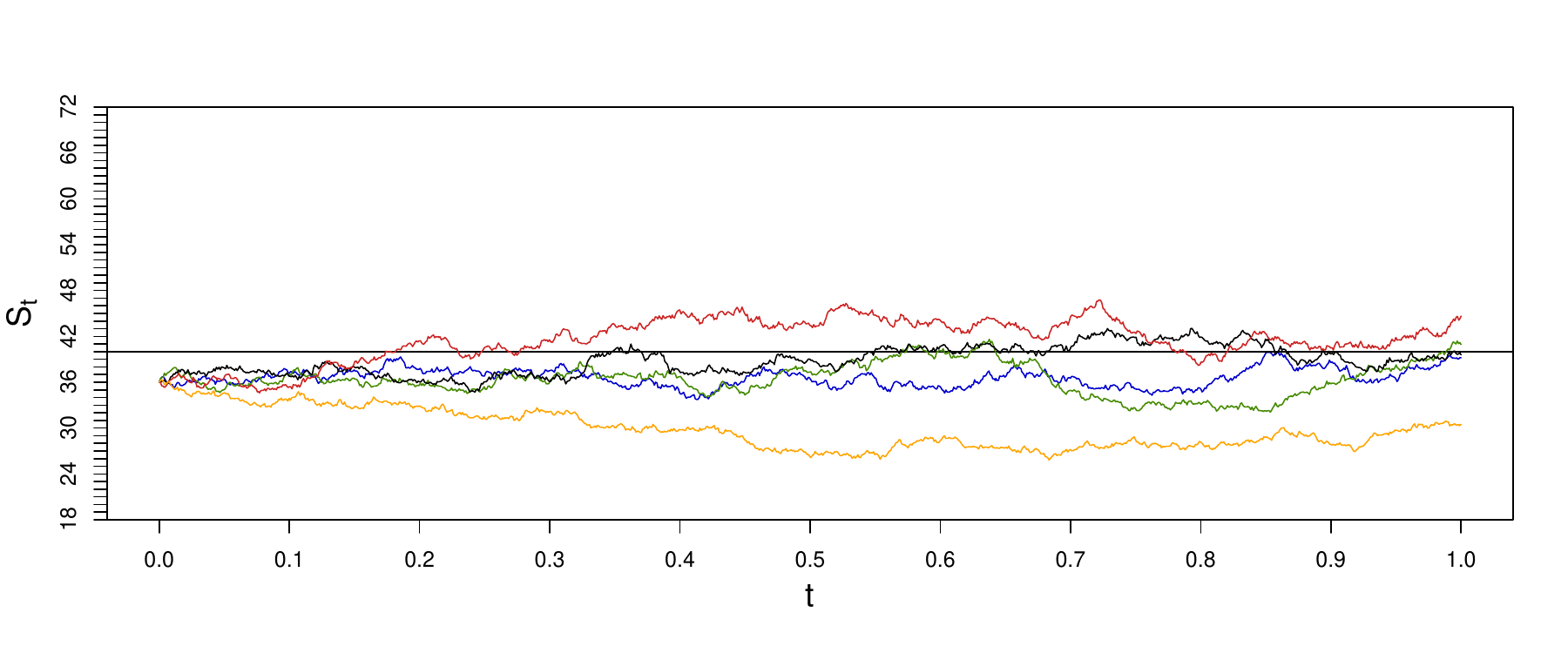}
        \caption{Asset price evolution for initial $S_0=36\,\$$ and $\lambda_0 = 10$ and $\eta=0.5$.}
    \end{subfigure}
    \vspace{0.5cm}

    \begin{subfigure}[t]{0.45\textwidth}
        \centering
        \includegraphics[width=\textwidth]{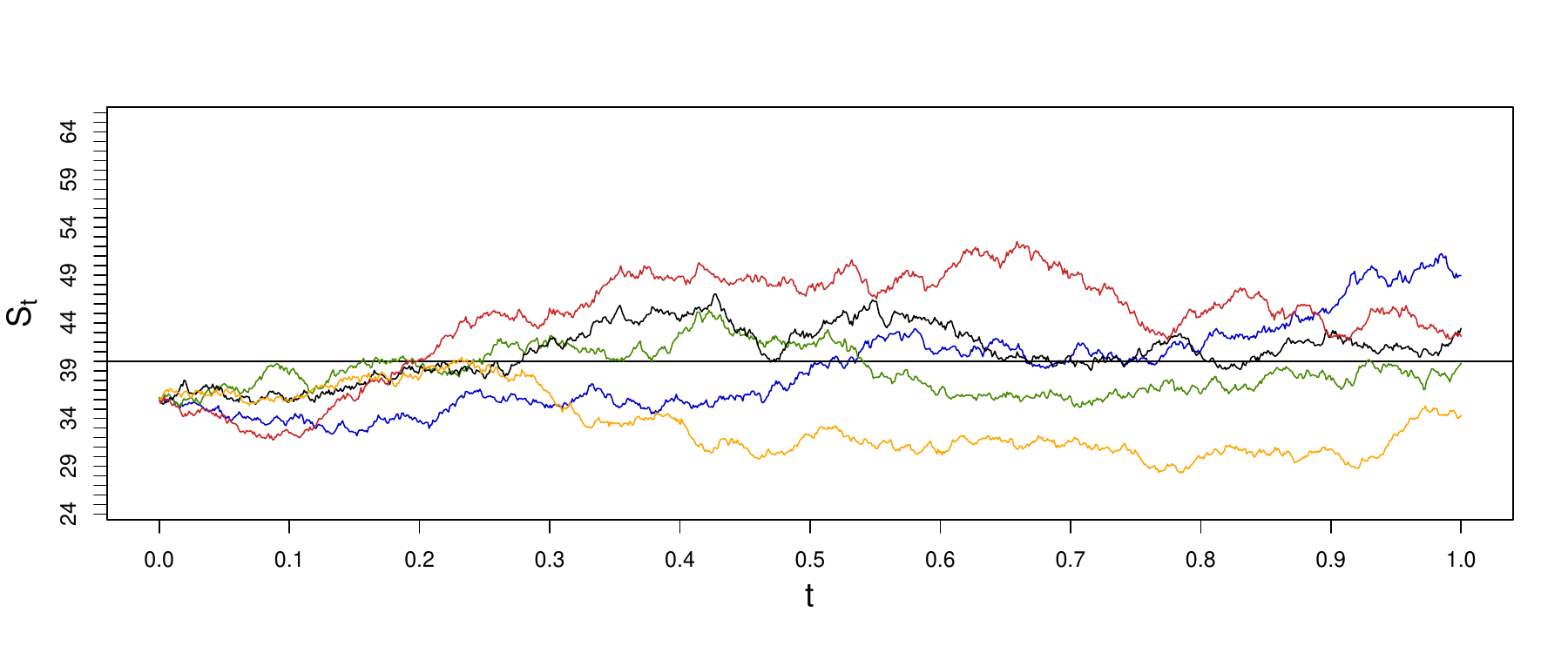}
        \caption{Asset price evolution for initial $S_0=36\,\$$ and $\lambda_0 = 1$ and $\eta=0.5$.}
    \end{subfigure}
    \hfill
    \begin{subfigure}[t]{0.45\textwidth}
        \centering
        \includegraphics[width=\textwidth]{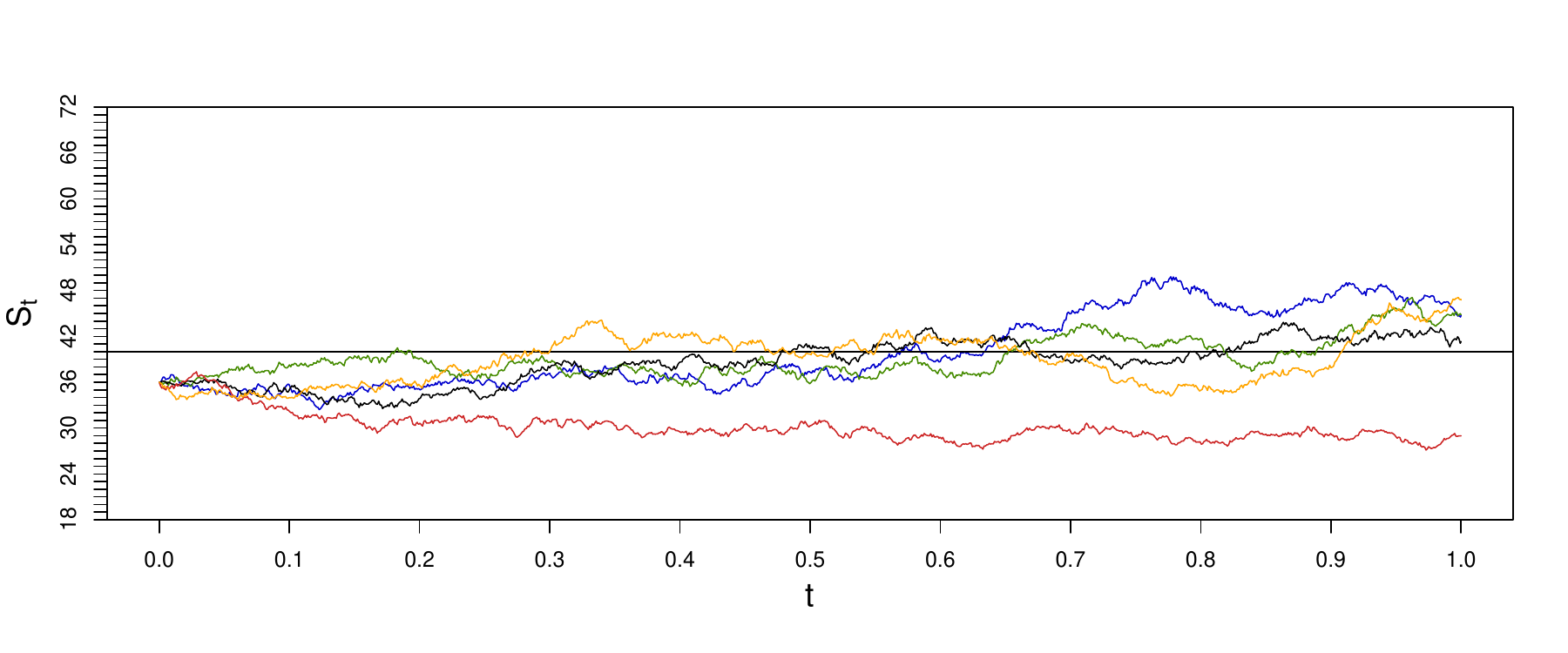}
        \caption{Asset price evolution for initial $S_0=36\,\$$ and $\lambda_0 = 1$ and $\eta=1$.}
    \end{subfigure}
    
    \caption{{\bf Scenario (D).} Numerical simulations of Equation \eqref{Micro_sys} with the parameters listed in Table \ref{parameter_mod1}. Here, we fix the initial value of the stock price $S_0=36\,\$$ for different values of $\eta$ and $\lambda_0$. The asset price evolution is shown after 1 year.}
\label{scenario_D}
\end{figure}
\noindent In the numerical test \textbf{(D)}, we examine the evolution of five asset price paths starting from an initial value $ S_0 = 36\,\$ $, with different settings for the jump intensity parameter $ \lambda_0 $ and the sensitivity parameter $ \eta $. The horizontal line indicates the strike price $ K = 40 \, \$ $. The results are shown in Figure \ref{scenario_D}. In subplot (a), where $ \lambda_0 = 10 $ and $ \eta = 0 $, the asset price paths appear relatively stable, with a slight upward trend for some paths, although most remain below the strike price $ K = 40 \, \$ $. The paths show limited fluctuations and, without sensitivity to deviations ($ \eta = 0 $), they do not appear to cluster tightly around the initial price $ S_0 = 36\,\$ $.\\
Moreover, in subplot (b), where $ \lambda_0 = 10 $ and $ \eta = 0.5 $, the asset price paths show a more clustered pattern around the initial value $ S_0 = 36\,\$ $. The introduction of some sensitivity ($ \eta = 0.5 $) tends to pull the paths closer to the initial value, although the overall spread seems to be slightly reduced compared to subplot (a). 
Furthermore, in subplot (c), where both $ \lambda_0 = 1 $ and $ \eta = 0.5 $), the asset price paths show what appears to be more significant upward movements compared to the previous plots. There appears to be an increasing divergence of the paths from the initial value $ S_0 = 36\,\$ $, with some paths approaching or even exceeding the strike price $ K = 40 \, \$ $. This observation suggests that reducing $ \lambda_0 $ while keeping $ \eta $ moderate may increase the variability and upward movement of the paths.\\
Finally, in subplot (d), the asset price paths appear to be more tightly clustered for both $ \lambda_0 = 1 $ and $ \eta = 1 $ than in subplot (c), although there is still considerable variability. The higher sensitivity parameter $ \eta = 1 $ seems to cluster the paths more tightly, reducing the overall dispersion. However, there appears to be a less pronounced upward movement compared to subplot (c), suggesting that the increased sensitivity may moderate the magnitude of the deviation from the initial price.
Overall, these observations suggest that $ \eta $ may control how tightly the paths are clustered around the initial value $ S_0 $, while $ \lambda_0 $ seems to influence the overall trend and the magnitude of the deviation from this initial value. 

\begin{figure}[H]
    \centering

     \begin{subfigure}[t]{0.45\textwidth}
        \centering
        \includegraphics[width=\textwidth]{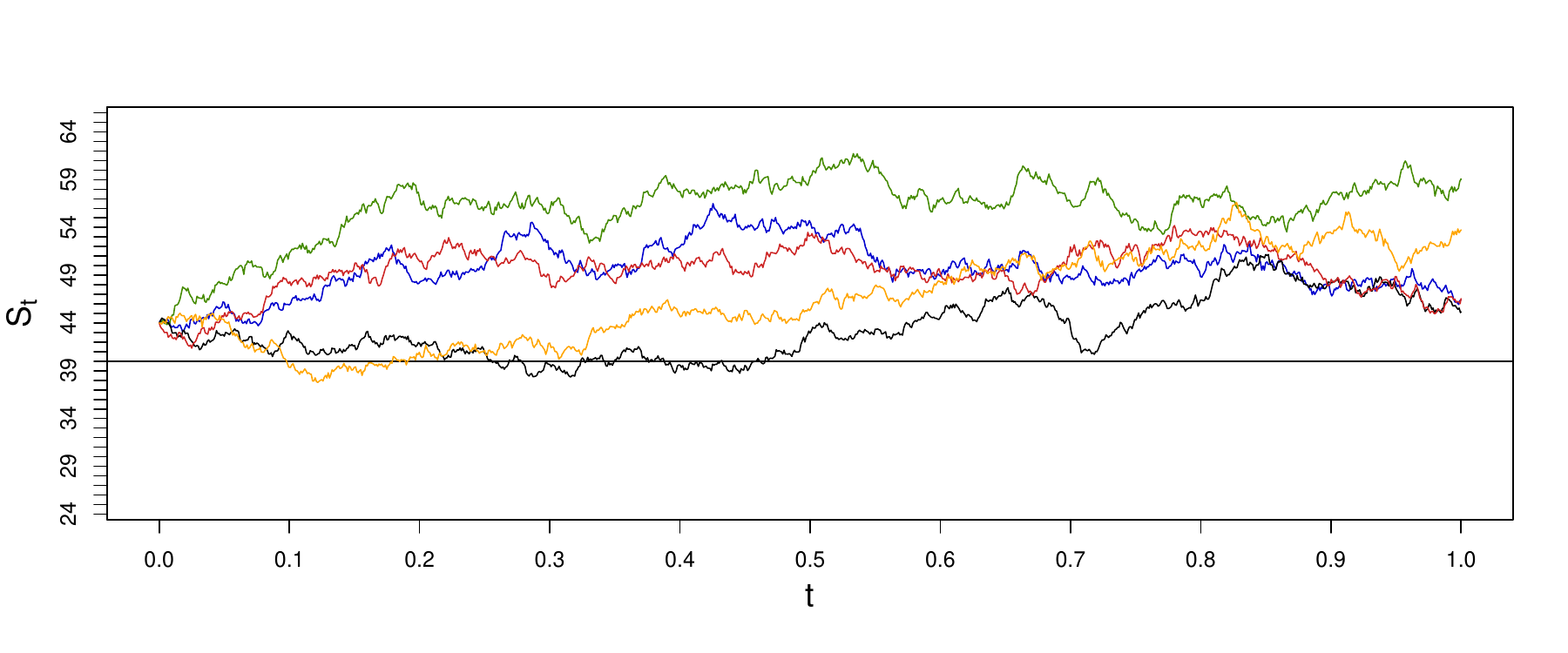}
        \caption{Asset price evolution for initial $S_0=44\,\$$ and $\lambda_0 = 10$ and $\eta=0$.}
    \end{subfigure}
    \hfill
    \begin{subfigure}[t]{0.45\textwidth}
        \centering
        \includegraphics[width=\textwidth]{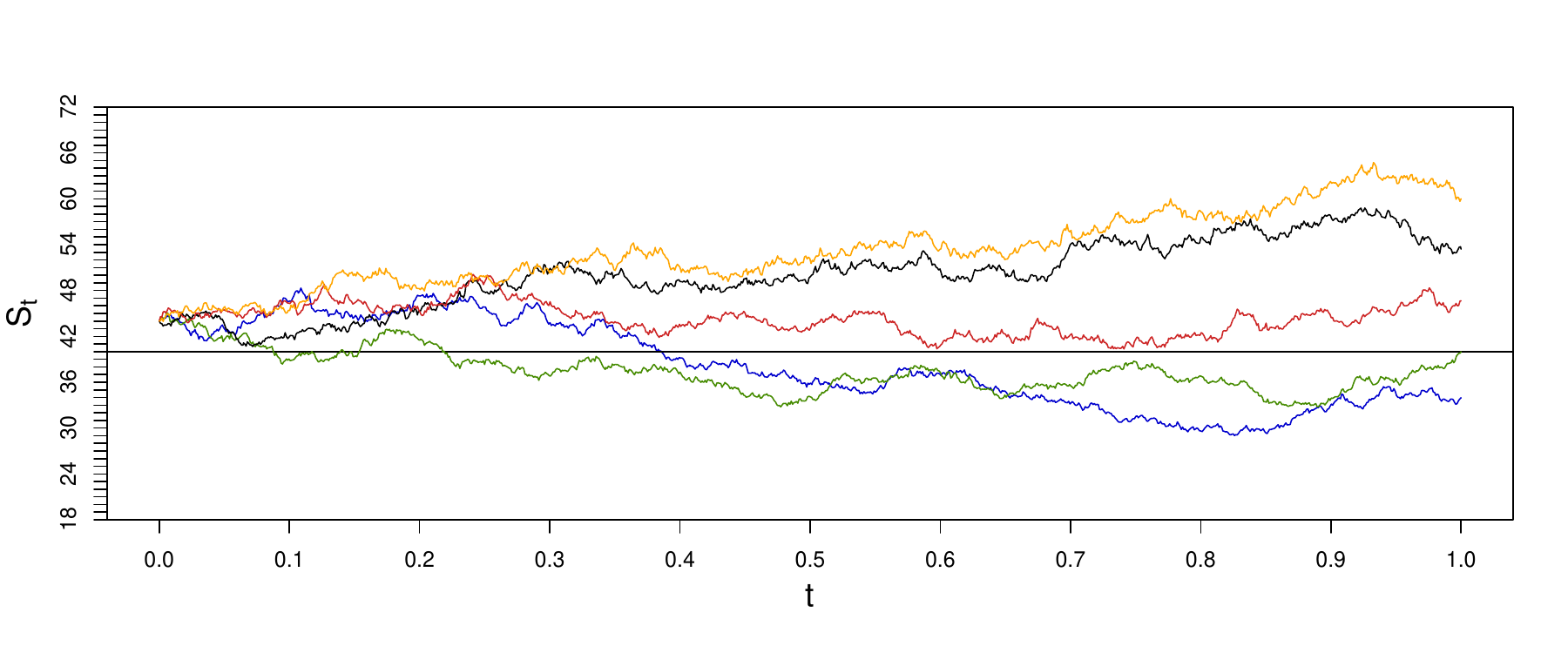}
        \caption{Asset price evolution for initial $S_0=44\,\$$ and $\lambda_0 = 10$ and $\eta=0.5$.}
    \end{subfigure}
    \vspace{0.5cm}

     \begin{subfigure}[t]{0.45\textwidth}
        \centering
        \includegraphics[width=\textwidth]{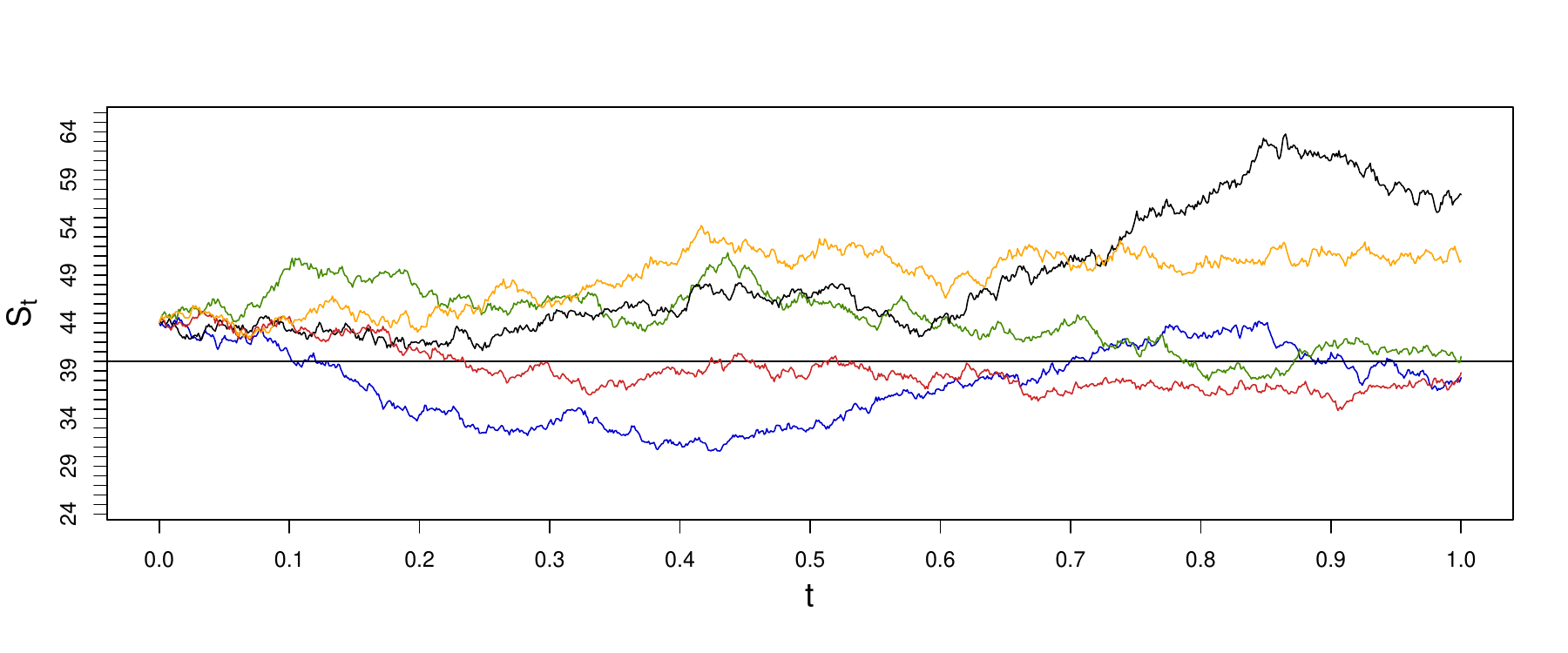}
        \caption{Asset price evolution for initial $S_0=44\,\$$ and $\lambda_0 = 1$ and $\eta=0.5$.}
    \end{subfigure}
    \hfill
    \begin{subfigure}[t]{0.45\textwidth}
        \centering
        \includegraphics[width=\textwidth]{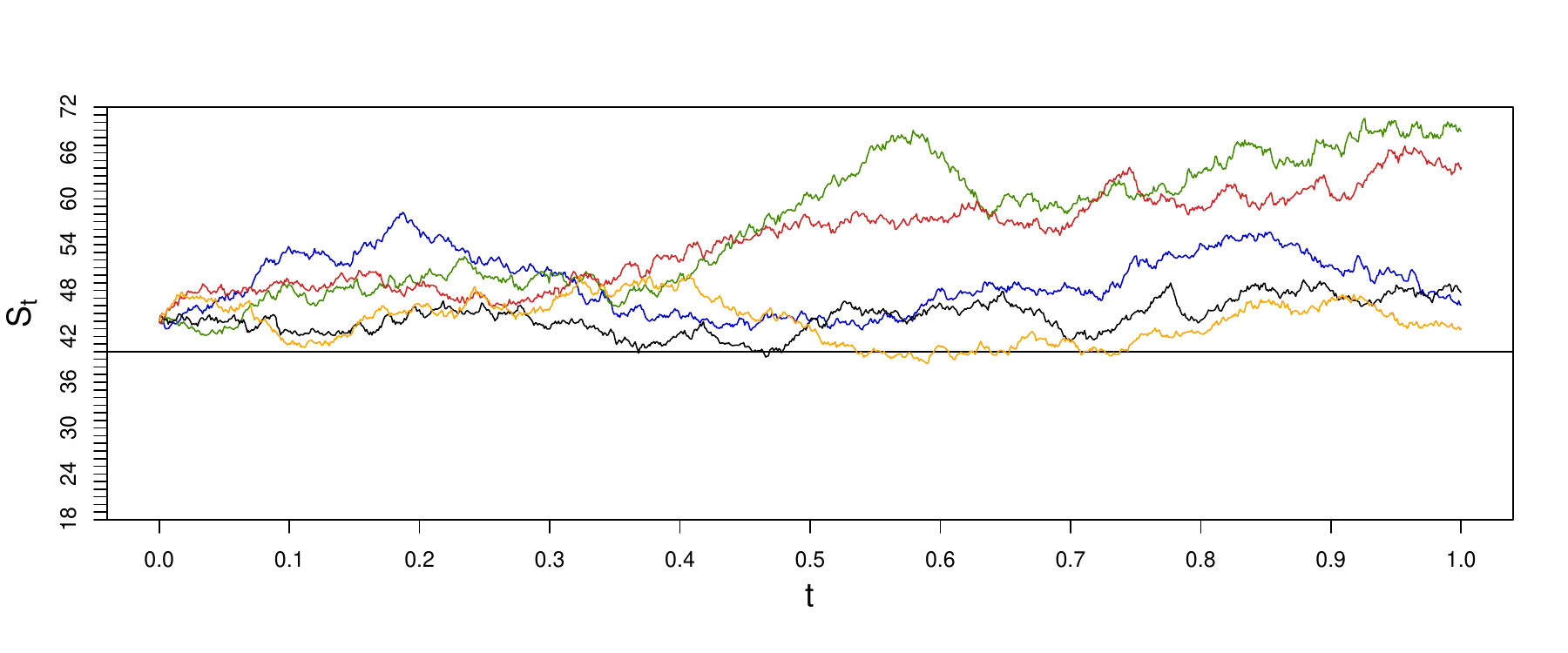}
        \caption{Asset price evolution for initial $S_0=44\,\$$ and $\lambda_0 = 1$ and $\eta=1$.}
    \end{subfigure}
    
    \caption{{\bf Scenario (E).} Numerical simulations of Equation \eqref{Micro_sys} with the parameters listed in Table \ref{parameter_mod1}. Here, we fix the initial value of the stock price $S_t=44\,\$$ for different values of $\eta$ and $\lambda_0$. The asset price evolution is shown after 1 year.}
\label{scenario_E}
\end{figure}

\begin{figure}[H]
    \centering
    \includegraphics[width=0.85\textwidth]{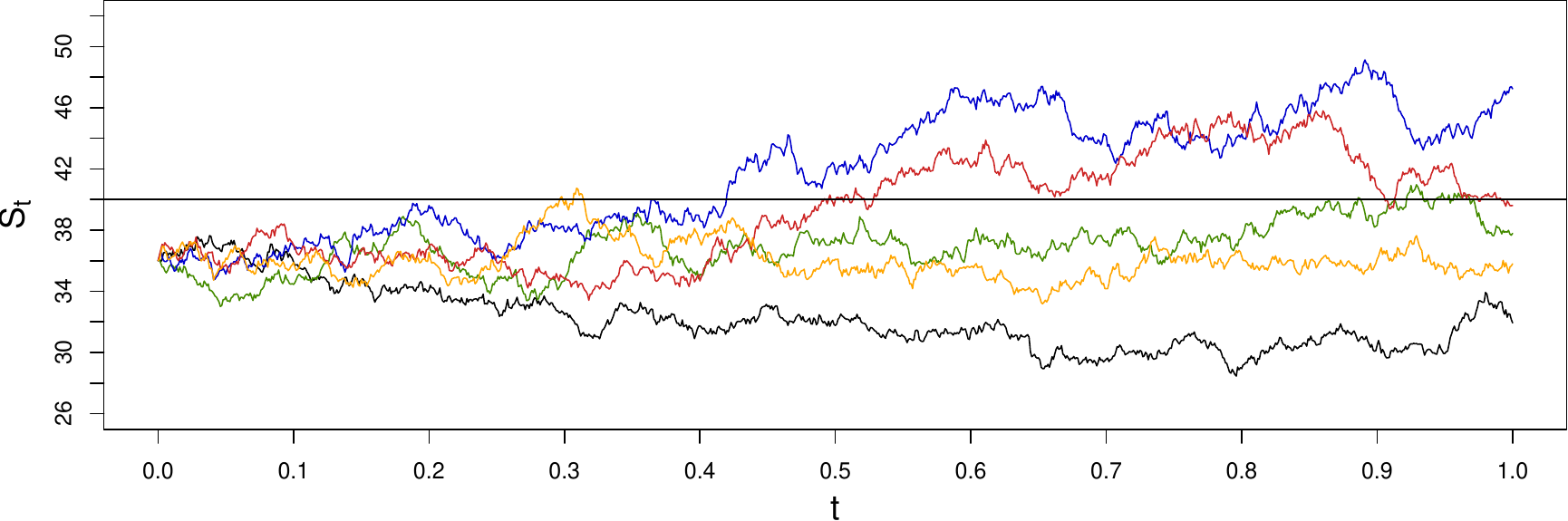}
    \caption{{\bf Model comparison.} Asset price evolution simulated using the GMB model with the same initial conditions and $\sigma$ values as in Scenario \textbf{(D)} and a constant drift. The evolution is shown after 1 year.}
    \label{BS_asset_price}
\end{figure}

\noindent In the numerical test \textbf{(E)}, we observe the evolution of five asset price paths, starting from an initial value of $ S_0 = 44\,\$ $, with different settings for the jump intensity parameter $ \lambda_0 $ and the sensitivity parameter $ \eta $. The horizontal line represents the strike price $ K = 40 \, \$ $, which is now lower than the initial asset price. The results are shown in Figure \ref{scenario_E}. In subplot (a), with $ \lambda_0 = 10 $ and $ \eta = 0 $, the asset price paths appear to be relatively stable with limited fluctuations, showing an upward trend away from the strike price $ K = 40 \, \$ $. The paths appear clustered and gradually move upwards, suggesting that without sensitivity to deviations ($ \eta = 0 $), the asset price paths are less likely to deviate significantly from each other or from the initial value $ S_0 $.\\
Introducing a moderate sensitivity, as shown in subplot (b), with $ \eta = 0.5 $ while keeping $ \lambda_0 = 10 $, the paths continue to show an upward trend, but with increased variability. The paths appear to cluster more closely around the initial value $ S_0 = 44\,\$ $ compared to subplot (a), but with more pronounced deviations, both upwards and downwards. This suggests that adding some sensitivity to the model allows the paths to respond more dynamically to market conditions, although they still tend to be mostly above the strike price. \\
However, when both $ \lambda_0 $ and $ \eta $ are reduced to 1 and 0.5 respectively, as in subplot (c), the paths appear to show more varied behaviour, with some paths tending to move downwards and closer to the strike price. The increased sensitivity seems to lead to a wider spread between the paths, reflecting a greater divergence in their trajectories. \\
This behaviour seems to be more pronounced in subplot (d), where $ \lambda_0 = 1 $ and $ \eta = 1 $. Here, the asset price paths show a high degree of variability, with significant upward movements in some paths, which reach much higher levels than in the other subplots. The increased sensitivity to deviations due to the higher value of $ \eta $ seems to result in more pronounced divergences between the paths, with some moving sharply upwards and others showing less drastic changes. In conclusion, the numerical test \textbf{(E)} suggests that $ \lambda_0 $ and $ \eta $ play a crucial role in shaping asset price trajectories, especially when the initial price starts above the strike price.\\
\noindent We now compare the asset price paths generated by the PDifMP model to those produced by the geometric Brownian motion (GBM) model used in the Longstaff-Schwartz algorithm. The GBM model, which assumes a constant drift and volatility, is a standard approach in financial modelling for simulating the evolution of asset prices. In contrast, our PDifMP model incorporates a time-varying drift and jump intensity, making it more responsive to market conditions. By comparing these two approaches, we aim to highlight the differences in how each model captures the dynamics of asset prices, particularly under varying market conditions.\\
Figure \ref{BS_asset_price} shows the evolution of asset prices over one year, simulated using the GMB model, which forms the basis of the LS algorithm. The initial conditions and volatility $\sigma$ in this simulation are the same as those used in the PDifMP scenario \textbf{(D)}. The horizontal line indicates the strike price $K = 40 \, \$$. Unlike the PDifMP model, which incorporates a time-varying drift, the GBM model assumes a fixed drift rate. This results in smoother and more predictable asset price paths in the GBM model, highlighting the limitations of the constant drift assumption compared to the more dynamic and responsive PDifMP model, which better captures the complexity of market dynamics.

\begin{figure}[H]
    \centering
    \includegraphics[width=\textwidth]{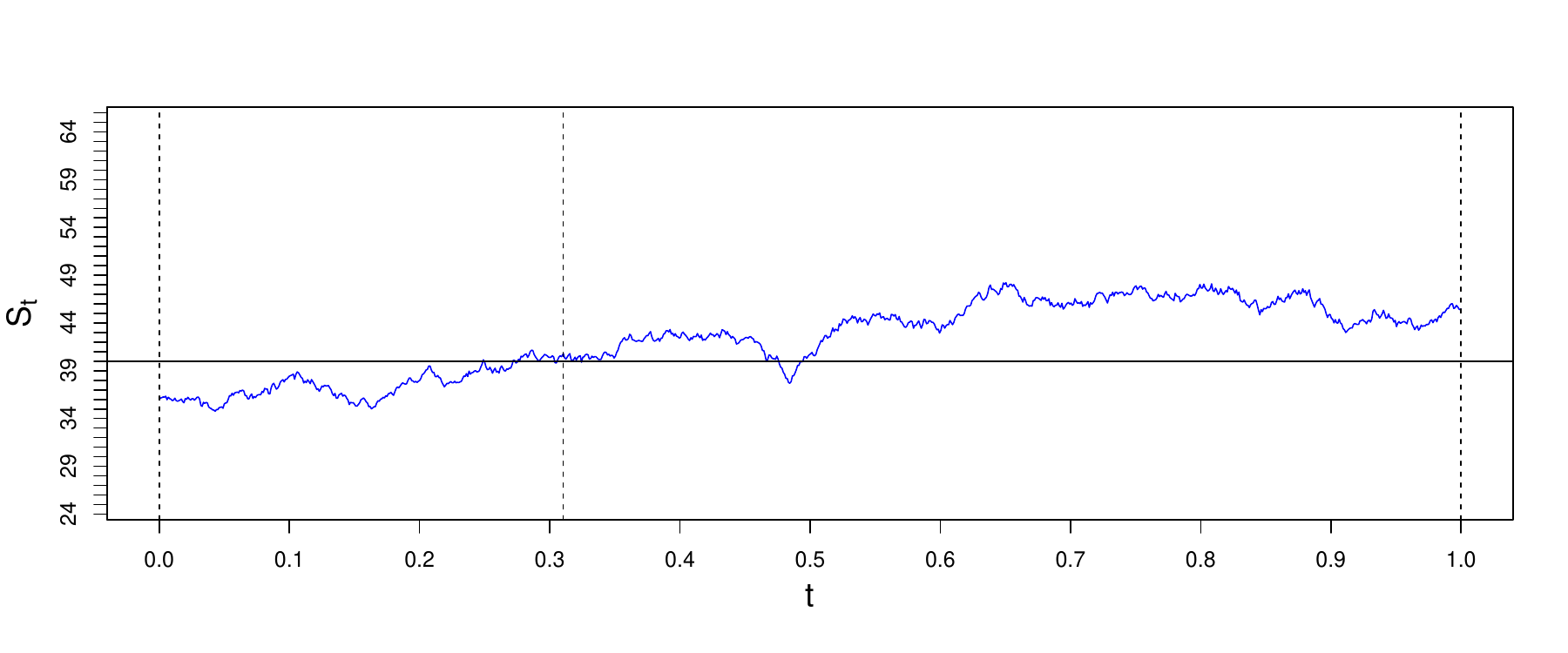}
\caption{{\bf Asset Price Evolution.} Asset price evolution computed using the PDifMP model with $\lambda_0 = 0.6$, $\eta = 0$, and $\alpha = 0.01$. The evolution is shown after 1 year.}    \label{JumpRateFromChapter5}
\end{figure}

\noindent To provide a context for the next section, we present Figure \ref{JumpRateFromChapter5}, which illustrates the evolution of asset prices computed using the PDifMP model with specific parameter values: $\lambda_0 = 0.6$, $\eta = 0$ and $\alpha = 0.01$. These values are representative of the parameters that will be examined in more detail in the following section. By examining this graph, we can observe the behaviour of the asset price under these conditions, including the number of jumps and the overall trend. This preliminary visualisation serves as a reference point for the more nuanced analysis to come, providing a glimpse of the dynamics that will be further explored with variations in these parameters. Moreover, the presence of fewer jumps in asset prices is consistent with real-world scenarios, where such jumps represent sudden deviations - either positive or negative - in asset prices. In reality, it is rare for a stock to experience significant trend changes more frequently than is the case here, making the output of this model a realistic approximation of typical market behaviour over the course of a year.

\section{Option pricing}
\label{section5}
Option pricing is a fundamental concept in financial mathematics, where the goal is to determine the fair value of an option — a financial derivative that grants the holder the right, but not the obligation, to buy (call option) or sell (put option) an underlying asset at a predetermined strike price $K$ before or at a specified expiration date $T$. The flexibility of American options, which can be exercised at any time up to expiration, adds complexity to their pricing compared to European options, which can only be exercised at maturity.\\
The payoff of a call option at time $t \in [0, T]$ is given by $\max(S_t - K, 0)$, where $S_t$ is the price of the underlying asset at time $t$. For a put option, the payoff is $\max(K - S_t, 0)$. When the payoff is positive, the option is said to be in the money.\\
This section compares three methods for pricing American options: the traditional Longstaff-Schwartz (LS) algorithm - which serves as a benchmark for the comparison - , a modified version of the LS algorithm using PDifMPs for asset price paths, and a novel approach that directly uses PDifMPs.
\subsection{The Longstaff-Schwartz algorithm}
The Longstaff-Schwartz algorithm, cf. \cite{LongSchw01}, is a widely recognised method for pricing American options that require the determination of an optimal exercise strategy. The key challenge in pricing these options is to decide whether to exercise or continue to hold the option at each possible time, based on a comparison between the immediate exercise payoff and the expected future payoff from continuing to hold.\\
To address this, the Longstaff-Schwartz algorithm employs Monte Carlo simulations to generate multiple possible paths for the price of the underlying asset over the life of the option. By analysing these paths, the algorithm estimates the conditional expectation of the continuation value at each potential exercise date. This estimation is achieved through regression, where the continuation value is modelled as a function of the underlying asset price. The fitted regression model provides a direct estimate of the conditional expectation, which is then compared with the immediate exercise payoff to determine the optimal exercise strategy. The procedure is iterative, working backwards from the expiration date of the option to the first exercise date, thus allowing the value of the American option to be calculated.\\
In more detail, the algorithm works as follows:  Consider a series of simulated paths for the underlying asset $S_t$, where $t \in [0, T]$. The option can be exercised at a series of discrete times $0 < t_1 \leq t_2 \leq \dots \leq t_K = T$. The algorithm proceeds iteratively, starting from the last exercise date at maturity $T$. At each time step $t_k$, the decision to exercise the option is based on whether the option is "in the money" (i.e. where the payoff is positive). For each path where the option is in the money, the immediate exercise payoff is compared to the conditional expectation of the continuation value. This continuation value is estimated by regressing the discounted value of future cash flows $Y$ on the current asset price $X$, using basis functions such as polynomials (e.g. linear, quadratic) or other functional forms such as Laguerre or Hermite polynomials.\\
Formally, let $X$ represent the prices of the underlying asset at time $t_k$, and let $Y$ denote the discounted value of subsequent option cash flows. The continuation value at time $t_k$ is estimated as the conditional expectation $E[Y \mid X]$, obtained from the regression model. If the immediate exercise payoff exceeds the estimated continuation value, the optimal strategy is to exercise the option, setting the cash flow at time $t_k$ to the immediate payoff and nullifying future cash flows for that path. Conversely, if the continuation value is higher, the cash flow at time $t_k$ is set to zero, and the path continues.\\
This process is repeated for each exercise date, moving backwards in time until the first exercise date is reached. The final value of the option is then computed as the average of the discounted cash flows across all simulated paths.\\
In the LS algorithm, asset price dynamics are usually modelled using a geometric Brownian motion (GMB) characterised by constant drift $\mu$ and volatility $\sigma$. The simulation of these asset price paths is outlined in Algorithm \ref{alg:SamplePaths}.
\begin{algorithm}[H]
\caption{Simulation of Sample Paths}
\begin{algorithmic}[1]
\Require Maturity $T$, step size $h$, initial value $S_0$, drift $\mu$, and volatility $\sigma$.
\State Set $N = T/h$.
\State Initialise a vector $path$ of length $N+1$.
\For{$i = 1$ to $N$}
    \State Draw $\xi_i \sim \mathcal{N}(0, h)$, a normally distributed random variable with mean 0 and variance $h$.
    \State Update the $i+1$-th entry of $path$ as $S_{i+1} = S_i \cdot \exp\left( (\mu - \frac{\sigma^2}{2}) \cdot h + \sigma \cdot \xi_i \right)$.
\EndFor
\State Output the vector $path$ containing the simulated asset prices.
\end{algorithmic}
\label{alg:SamplePaths}
\end{algorithm}
\noindent Further, we detail below the implementation of the Longstaff-Schwartz algorithm specifically for put and call options. This algorithm uses the simulated asset price paths generated in the previous step to determine the option price, as outlined in Algorithm \ref{alg:LS}.
\begin{table}[H]
\begin{center}
\begin{tabular}{c|c|c} 
    \hline  
    \rule{0pt}{3ex}Parameter & Description & Value (unit) \\[1ex]
    \hline
    \rule{0pt}{2ex} $S_0$ & initial stock price    & $36-44\,\$$ \\[1.5ex]
    \rule{0pt}{2ex} $K$ & strike price & $40\$$\\[1.5ex]
    \rule{0pt}{2ex} $\mu$ & drift rate  & $0.06$   \\[1.5ex]
    \rule{0pt}{2ex} $\sigma$ & volatility  & $0.2$   \\[1.5ex]
    \rule{0pt}{2ex} $h$ & step size  & $0.02$   \\[1.5ex]
    \rule{0pt}{2ex} $T$ & maturity  & $1$ year  \\[1.5ex]
    \rule{0pt}{2ex} $M$ & number of asset price paths  & $10000$   \\[1.5ex]
    \rule{0pt}{2ex} $M_E$ & number of exercise points  & $50$   \\[1.5ex]
    \hline
\end{tabular}
\end{center}
\vspace{0.3cm}
\caption{\textbf{Model parameters.} Parameters used in the LS algorithm.}
\label{parameter_LS}
\end{table}
\noindent The simulation of sample paths and implementation of the Longstaff-Schwartz algorithm will use the parameters in Table \ref{parameter_LS} unless otherwise specified.
\begin{algorithm}
\caption{Longstaff-Schwartz algorithm}
\begin{algorithmic}[1]
\Require Maturity $T$. Matrix $paths$ with $M$ rows (each representing a simulated asset price path) and $M_E$ columns (each representing an exercise time), and the strike price $K$.
\State Initialise a matrix $cash\_flows$ of size $M \times M_E$ with zeros.
\State Set the last column of $cash\_flows$ to the inner value for each path, i.e., ${\max(K - paths[:, M_E], 0)}$.
\For{$t = M_E-1$ to $2$ (step backwards)}
    \State Set the vector $X$ of asset prices at time $t$, i.e., $X = paths[:, t]$.
    \State Identify the paths where the option is in the money, i.e., $in\_the\_money = \{i : X[i] < K\}$.
    \State Extract the corresponding $X$ values, $X\_in\_the\_money$, and the discounted future cash flows $Y\_in\_the\_money$ for these paths.
    \State Perform a regression of $Y\_in\_the\_money$ on a constant, $X\_in\_the\_money$, and $X\_in\_the\_money^2$ to estimate the continuation value.
    \State Calculate the inner value vector $inner\_value = K - X\_in\_the\_money$.
    \For{$j = 1$ to $|in\_the\_money|$}
    \For{$i$ in $in\_the\_money$}
        \If{$inner\_value[j] > continuation\_values[j]$}
            \State Set $cash\_flows[i,t] = inner\_value[j]$.
            \State Set all future cash flows for this path to 0, i.e., $cash\_flows[i,(t+1):M_E] = 0$.
        \Else
            \State Set $cash\_flows[i,t] = 0$.
        \EndIf
        \State  $j=j+1$.
    \EndFor
\EndFor

\EndFor
\State Compute the option price as the average of the discounted cash flows.
\end{algorithmic}
\label{alg:LS}
\end{algorithm}

\subsection{Longstaff-Schwartz algorithm with PDifMP-driven paths}
We present here a modified version of the LS algorithm that incorporates PDifMPs into the asset price simulation, which we refer to as LS+PDifMP. Unlike the GBM model, PDifMPs combine continuous dynamics with discrete jumps, allowing them to more accurately model both gradual trends and sudden market shifts. By using PDifMP-generated paths instead of GBM, the LS+PDifMP approach can better reflect the time-varying nature of market conditions, especially in scenarios with significant volatility.\\
In this approach, the LS algorithm operates on asset price trajectories generated by PDifMPs that are influenced by the deviation of the asset from its initial value. This change is expected to increase the flexibility and accuracy of the model, allowing it to better capture the dynamic nature of market conditions. As a result, the LS+PDifMP approach has the potential to be a more effective tool for pricing American options, particularly in volatile markets. The simulation process is conducted by combining Algorithm \ref{alg:assetPricePDifMP}, which generates PDifMP paths, with Algorithm \ref{alg:LS}, which applies the LS algorithm to these paths. Unless otherwise specified, the parameters listed in Table \ref{parameter_LSandPDifMP} will be used for these simulations.
\begin{table}[H]
\begin{center}
\begin{tabular}{c|c|c} 
    \hline  
    \rule{0pt}{3ex}Parameter & Description & Value (unit) \\[1ex]
    \hline
    \rule{0pt}{2ex} $S_0$ & initial stock price    & $36-44\,\$$ \\[1.5ex]
    \rule{0pt}{2ex} $\lambda_0$  & minimum frequency of fluctuation & $0.4-10$ \\[1.5ex] 
    \rule{0pt}{2ex} $\eta$  & sensitivity of fluctuation in $\lambda(U_t)$ &$0-0.6$\\[1.5ex]
    \rule{0pt}{2ex} $K$ & strike price & $40\,\$$\\[1.5ex]
    \rule{0pt}{2ex} $\mu_0$ & initial drift  & $0.06$   \\[1.5ex]
    \rule{0pt}{2ex} $\sigma$ & stochastic parameter  & $0.2$   \\[1.5ex]
    \rule{0pt}{2ex} $\alpha$ & scaling factor & $0.01$   \\[1.5ex]
    \rule{0pt}{2ex} $b$ & scaling factor & $0.01$   \\[1.5ex]
    \rule{0pt}{2ex} $T$ & maturity  & $1$   \\[1.5ex]
    \rule{0pt}{2ex} $h$ & step size  & $10^{-3}$   \\[1.5ex]
    \rule{0pt}{2ex} $r$ & interest rate  & $0.06$   \\[1.5ex]
    \rule{0pt}{2ex} $M$ & number of paths of the asset price  & $10 000$   \\[1.5ex]
    \rule{0pt}{2ex} $M_E$ & number of exercise points  & $50$   \\[1.5ex]
    \hline
\end{tabular}
\end{center}
\vspace{0.3cm}
\caption{\textbf{Model parameters.} Parameters used in the LS+PDifMP algorithm.}
\label{parameter_LSandPDifMP}
\end{table}

\subsection{PDifMP-based option pricing method}
As an alternative to the regression-based approach of the Longstaff-Schwartz algorithm, we propose a new approach that directly employs PDifMPs to price American options. This method leverages the inherent properties of PDifMPs, in particular the random jump times, as potential exercise points for the option. In doing so, it simplifies the computational process and provides a direct mechanism for determining the option price, eliminating the need for backward iteration and regression to estimate continuation values.\\
The core idea of this approach is to model the asset price as a PDifMP, where the price follows a continuous path with random jumps occurring at discrete times.\\
These jumps, determined by the intensity function $\lambda(U_t)$ as defined in Equation \eqref{jump_Rate}, are treated as potential option exercise times. The flexibility of the PDifMP framework allows us to vary the number and distribution of jump times by adjusting the jump rate function $\lambda(U_t)$. This adaptability allows the model to approximate the results obtained by the Longstaff-Schwartz algorithm under different market conditions.\\
In this method, the intrinsic value of the option is computed at each jump time $T_n$. For a put option, this intrinsic value is given by $\max(K - S_{T_n}, 0)$, where $K$ is the strike price and $S_{T_n}$ is the asset price at time $T_n$. For a call option, the intrinsic value is $\max(S_{T_n} - K, 0)$. These intrinsic values are then discounted to the present value using the discount factor $e^{-rT_n}$, where $r$ represents the risk-free interest rate. The final option price for each path is determined by taking the maximum of these discounted intrinsic values across all jump times along that specific path. Given $M$ simulated paths, we let $ P_m $ denote the maximum value of the discounted intrinsic values across all possible exercise times for the put option along the $ m$-th path, where $m=1,\ldots, M$.  This value $ P_m $ is given by
\begin{equation*}
    P_m = \max_{n = 1,\dots,N} \left[ e^{-rT_n} \max(K - S_{T_n}, 0) \right],
\end{equation*}
where $ e^{-rT_n} $ is the discount factor applied to the intrinsic value $ \max(K - S_{T_n}, 0) $ at each jump time $ T_n $, and $ N $ is the total number of jumps along the $m$-th path.\\
Similarly, for the call option, let $ C_m $ represents the maximum value of the discounted intrinsic values across all possible exercise times for the call option along the path $ m $, such that 
\begin{equation*}
    C_m = \max_{n = 1,\dots,N} \left[ e^{-rT_n} \max(S_{T_n} - K, 0) \right].
\end{equation*}
The final option price is then obtained by averaging these values across all simulated paths:

\begin{equation*}
	\text{Option Price}= \left\{
	\begin{array}{ll}
		\frac{1}{M}\sum_{m=1}^{M} P_m \qquad\qquad  \text{for the put option},\\[0.2cm]
		\frac{1}{M}\sum_{m=1}^{M} C_m \qquad\qquad  \text{for the call option}.
	\end{array}
	\right. 
\end{equation*}

\begin{algorithm}[H]
\caption{Option Pricing with PDifMP}
\begin{algorithmic}[1]
\Require Maturity, $T$, strike price $K$, number of paths $M$.
\State Initialise an empty vector $optionValues$ to store the option values for each path.
\For{$i = 1$ to $M$}
    \State Use Algorithm \ref{alg:assetPricePDifMP} to generate a path $S$ of the asset price with corresponding jump $times$.
    \State Initialise an empty vector $discountedInnerValues$ to store the discounted intrinsic values.
    \For{$t$ in $times$}
        \State Calculate the discounted intrinsic value at time $t$ and add it to $discountedInnerValues$:
        \State \quad For put option: $e^{-rt} \max(K - S_t, 0)$
        \State \quad For call option: $e^{-rt} \max(S_t - K, 0)$
    \EndFor
    \State Add the maximum of $discountedInnerValues$ to $optionValues$.
\EndFor
\State Output the average of $optionValues$.
\end{algorithmic}
\label{alg:optionPricePDifMP}
\end{algorithm}

\noindent The primary advantage of this method lies in its ability to avoid the complexities of backward iteration and regression that are inherent in the LS algorithm. By directly leveraging the jump times in PDifMPs, this approach offers a more straightforward and potentially more efficient way to price American options. However, the effectiveness of this method is heavily influenced by the choice of the jump rate function $\lambda(U_t)$. Low jump rates result in fewer exercise opportunities, which can lower the option price, while higher jump rates provide more frequent exercise opportunities, potentially raising the option price. Therefore, carefully tuning $\lambda(U_t)$ to match specific market conditions is crucial for optimising the performance of the model.

\subsection{Comparative analysis of Put option pricing methods}
In this section, we compare the performance of the three pricing methods—Longstaff-Schwartz (LS), Longstaff-Schwartz with PDifMP (LS+PDifMP), and the direct PDifMP method—specifically for put options.  To ensure a fair and comprehensive comparison, we apply each method under the same set of initial conditions and parameters, as outlined in Table \ref{parameter_LS} and \ref{parameter_LSandPDifMP}, respectively. The asset price paths are simulated under identical market assumptions, with variations introduced only in the methods themselves. The strike price $K$ and the initial asset price $S_0$ are fixed, while the minimum frequency $\lambda_0$ and the sensitivity parameter $\eta$ are varied to explore different market conditions. The risk-free interest rate $r$ is also held constant across all simulations. We recall the jump rate function defined as in Equation \eqref{jump_Rate} (Section \ref{Charct_of_the_drift}). In this work, we set $\beta = 0$ in Equation \eqref{jump_Rate} to simplify the jump intensity function, making it directly responsive to any deviation of the asset price $S_t$ from the reference point $\delta$. This approach is suitable for scenarios where even small price movements relative to $\delta$ are significant. However, we recognise that in other market conditions, a non-zero $\beta$ could introduce a buffer zone around $\delta$, allowing the model to maintain a baseline jump rate $\lambda_0$ and avoid overreacting to minor fluctuations that are not critical \\
Therefore, the jump rate function in this context is given by
\begin{equation*}
     \lambda(U_t) := \lambda_0 + \eta \max\left(0, (|S_t - \delta|)\right),   
\end{equation*}
and the location parameter of the Laplace distribution given by
\begin{equation*}
   a(S_t) = \mu_0 + \alpha(S_t - \delta).
\end{equation*}
To thoroughly evaluate their impact on option pricing, we will split the comparison into two parts. First, we will consider the case where the jump rate function and the location parameter depend on the distance between the current asset price $S_t$ and the initial asset price $S_0$, i.e. $\delta=S_0$. In this scenario, the jump rate increases as the asset price deviates from its initial value, capturing the idea that larger movements away from the starting price could indicate greater market volatility and, therefore, a higher likelihood of jumps.\\
Next, we will analyse the scenario where the jump rate and the location parameter are based on the distance between the current asset price $S_t$ and the strike price $K$, i.e. $\delta=K$. In this case, the jump rate function is sensitive to the moneyness of the option, meaning that the probability of jumps changes as the asset price approaches or moves away from the strike price. This reflects market dynamics where significant deviations from the strike price can lead to abrupt changes in the value of the option.
Under these two scenarios, we will run different sets of simulations to assess how each jump rate function influences the pricing of put options across various market conditions.\\
As a first step in our analysis, we fix the values of the parameters as follows: $\alpha = 10^{-6}$, $\lambda_0 = 5$, and $\eta = 0.5$. The remaining parameters are taken according to Table \ref{parameter_LSandPDifMP}. We then compare the classic LS algorithm with the modified LS+PDifMP method for different initial asset prices $S_0$. For each comparison, we consider two scenarios: one where the parameter $\delta$ is set equal to the strike price $K$ and another where $\delta$ is set equal to the initial asset price $S_0$. The results of this comparison are presented in Table \ref{tab:LSandPDifMP_St-x}:

\begin{table}[H]
\begin{center}
\begin{tabular}{|c|c|c|c|}
    \hline 
    $S_0$ & $\delta$ & LS Classic & LS+PDifMP \\  \hline \hline

    36 & 40 ($K$) & 4.472 & 5.450 \\  \hline 
    36 & 36 ($S_0$) & 4.472 & 4.458 \\  \hline \hline
    
    38 & 40 ($K$) & 3.244 & 4.091 \\ \hline 
    38 & 38 ($S_0$) & 3.244 & 3.264 \\ \hline \hline
    
    40 & 40 ($S_0$) & 2.313 & 2.358 \\  \hline \hline

    42 & 40 ($K$) & 1.617 & 1.599  \\  \hline 
    42 & 42 ($S_0$) & 1.617 & 1.681 \\  \hline \hline

    44 & 40 ($K$) & 1.118 & 0.933 \\ \hline
    44 & 44 ($S_0$) & 1.118 & 1.166 \\ \hline
\end{tabular}    
\end{center}
\vspace{0.3cm}
 \caption{\textbf{Comparison of Put Option Prices Using LS and LS+PDifMP.} The table presents put option prices calculated using the classic LS algorithm and the modified LS+PDifMP method across various initial asset prices $S_0$. The values used for the parameters are $\alpha = 10^{-6}$, $\lambda_0 = 5$, $\eta = 0.5$.}
\label{tab:LSandPDifMP_St-x}
\end{table}
\noindent The results in Table \ref{tab:LSandPDifMP_St-x} highlight notable differences in the calculated put option prices between the classic LS algorithm and the LS+PDifMP method for various initial asset prices $S_0$ and scenarios where $\delta$ is set to either the strike price $K$ or the initial asset price $S_0$. When $\delta$ is set to $K$, the LS+PDifMP method generally produces higher option prices when $S_0$ is significantly below the strike price. This result suggests that the LS+PDifMP method is more sensitive to the likelihood of continued downward movements in the asset price, which increases the value of the put option under these conditions. However, as $S_0$ approaches or exceeds $K$, the differences diminish or even reverse, suggesting that the LS+PDifMP method adjusts for lower volatility or potential upward movements as the asset price approaches or exceeds the strike price. Conversely, when $\delta$ is set to $S_0$, the LS+PDifMP method tends to produce prices very close to or slightly above those of the LS classic method, especially when $S_0$ is close to the strike price. This suggests that the method is sensitive to deviations from the initial asset price and provides a nuanced adjustment of the option price based on the expected volatility around $S_0$. Overall, the results suggest that the LS+PDifMP method introduces valuable flexibility by varying the jump rate based on proximity to either the strike price or the initial asset price, potentially offering a more accurate reflection of market dynamics, although the effectiveness of this approach is highly dependent on the choice of $\delta$ and the specific market conditions being modelled.\\
Next, we fix $\delta$ to $S_0$ and conduct a series of experiments to further explore the performance of the LS+PDifMP method and PDifMP method compared to the classic LS algorithm under various conditions.
\begin{itemize}
    \item \textbf{Experiment A.1}: We set $\alpha = 0.01$ and $\eta = 0$, and then compare the LS algorithm with the LS+PDifMP and the PDifMP method across different values of $S_0$ and $\lambda_0$.
    \item \textbf{Experiment A.2}: We set $\alpha = -0.01$ and $\eta = 0$, and similarly compare the LS algorithm with the LS+PDifMP and the PDifMP method for different values of $S_0$ and $\lambda_0$.
    \item \textbf{Experiment A.3}: We fix $\eta = 0.005$ and $\lambda_0$, and then compare the three methods to study the effect of varying $\alpha$.
\end{itemize}

\noindent In Experiment \textbf{A.1}, we set $\alpha = 0.01$ and $\eta = 0$, and compared the performance of the LS algorithm, the LS+PDifMP method, and the direct PDifMP approach across various initial asset prices $S_0$ and different values of the jump rate parameter $\lambda_0$. The results are summarised in Table \ref{tab:alpha0.01}. 

\begin{table}[H]
\begin{center}
\begin{tabular}{|c|c|c|c|c|}
    \hline 
    $S_0$ & $\lambda_0$  & Longstaff-Schwartz & LS + PDifMP & PDifMP  \\  \hline \hline
    36 & 0.4  & 4.472 & 4.443 & 4.179\\ \hline
    36 & 0.6 &  4.472 & 4.429 & 4.488\\ \hline
    36 & 0.8 &  4.472 & 4.418 & 4.591\\ \hline \hline

    38 & 0.4 & 3.244 & 3.212 & 3.145\\ \hline
    38 & 0.6 & 3.244 & 3.242 & 3.329\\ \hline
    38 & 0.8 & 3.244 & 3.164 & 3.397\\ \hline \hline

    40 & 0.4 & 2.313 & 2.290 & 2.309\\ \hline
    40 & 0.6 & 2.313 & 2.290 & 2.383\\ \hline
    40 & 0.8 & 2.313 & 2.300 & 2.455\\ \hline \hline

    42 & 0.4 & 1.617 & 1.620 & 1.558\\ \hline
    42 & 0.6 & 1.617 & 1.628 & 1.605\\ \hline
    42 & 0.8 & 1.617 & 1.552 & 1.709\\ \hline \hline

    44 & 0.4 & 1.118 & 1.149 & 1.116*\\ \hline
    44 & 0.6 & 1.118 & 1.118 & 1.130*\\ \hline
    44 & 0.8 & 1.118 & 1.153 & 1.181*\\ \hline 
\end{tabular}
\end{center}
\vspace{0.3cm}
\caption{\textbf{Experiment A.1.} Put Option prices calculated using the LS, LS+PDifMP, and PDifMP methods with $\eta = 0$ and $\alpha = 0.01$, for different values of $\lambda_0$ and $S_0$.}
\label{tab:alpha0.01}
\end{table}
The results in Table \ref{tab:alpha0.01} show that the put option prices obtained using the LS, LS+PDifMP and PDifMP methods are quite close, with only slight variations for different initial asset prices $S_0$ and jump rates $\lambda_0$. The LS+PDifMP method is generally close to the classical LS algorithm, with only minor adjustments. The PDifMP method shows slightly more variation, especially as $\lambda_0$ increases, but the differences remain modest. Some PDifMP prices, marked with an asterisk, are based on simulations where the asset price exceeded three times its initial value, which may be unrealistic. Overall, the results suggest that the inclusion of PDifMPs leads to subtle changes in option pricing under the given conditions.
\begin{table}[H]
\begin{center}
\begin{tabular}{|c|c|c|c|c|}
    \hline 
    $S_0$ & $\lambda_0$  & Longstaff-Schwartz & LS + PDifMP & PDifMP  \\  \hline \hline
    36 & 0.4 & 4.472 & 4.473 & 4.271 \\ \hline
    36 & 0.6 & 4.472 & 4.535 & 4.451 \\ \hline
    36 & 0.8 & 4.472 & 4.492 & 4.612 \\ \hline \hline

    38 & 0.4 & 3.244 & 3.257 & 3.113 \\ \hline
    38 & 0.6 & 3.244 & 3.259 & 3.268 \\ \hline
    38 & 0.8 & 3.244 & 3.283 & 3.393 \\ \hline \hline

    40 & 0.4 & 2.313 & 2.289 & 2.305 \\ \hline
    40 & 0.6 & 2.313 & 2.259 & 2.327 \\ \hline
    40 & 0.8 & 2.313 & 2.308 & 2.365 \\ \hline \hline

    42 & 0.4 & 1.617 & 1.642 & 1.534 \\ \hline
    42 & 0.6 & 1.617 & 1.622 & 1.626\\ \hline
    42 & 0.8 & 1.617 & 1.626 & 1.628 \\ \hline \hline

    44 & 0.4 & 1.118 & 1.103 & 1.065\\ \hline
    44 & 0.6 & 1.118 & 1.091 & 1.058\\ \hline
    44 & 0.8 & 1.118 & 1.083 & 1.071\\ \hline 
\end{tabular}    
\end{center}
\vspace{0.3cm}
\caption{\textbf{Experiment A.2.} Put Option prices calculated using the LS, LS+PDifMP, and PDifMP methods with $\eta = 0$ and $\alpha = -0.01$, for different values of $\lambda_0$ and $S_0$.}
\label{tab:AlphaMinus0.01}
\end{table}
\noindent The results in Table \ref{tab:AlphaMinus0.01} show the calculated put option prices using the classic Longstaff-Schwartz (LS) algorithm, the LS+PDifMP method, and the PDifMP method for various initial asset prices $S_0$ and base jump rates $\lambda_0$, with parameters $\eta = 0$ and $\alpha = -0.01$.\\
Compared to the previous case where $\alpha = 0.01$, the option prices in this scenario are generally similar, with only minor differences observed between the methods. The LS+PDifMP method continues to produce prices that are very close to those obtained with the LS algorithm, suggesting that the negative value of $\alpha$ does not significantly alter the pricing dynamics.  The PDifMP method shows slightly more variation than the LS and LS+PDifMP methods, especially as $\lambda_0$ increases. However, the differences remain small. In particular, in this case, there are no option prices based on simulations where the asset price exceeded three times its initial value, which means that the potential for unrealistic pricing is not a concern here, unlike in the previous case.

\begin{table}[H]
\begin{center}
\begin{tabular}{|c|c|c|c|c|}
    \hline 
    $S_0$ & $\alpha$  & Longstaff-Schwartz & LS + PDifMP & PDifMP  \\  \hline \hline
    36 & 0.01  & 4.472 & 4.386 & 4.499 \\ \hline
    36 & -0.01 & 4.472 & 4.449 & 4.540 \\ \hline \hline

    38 & 0.01  & 3.244 & 3.172 & 3.342 \\ \hline
    38 & -0.01 & 3.244 & 3.277 & 3.379 \\ \hline \hline

    40 & 0.01  & 2.313 & 2.296 & 2.411 \\ \hline
    40 & -0.01 & 2.313 & 2.293 & 2.356 \\ \hline \hline

    42 & 0.01  & 1.617 & 1.623 & 1.691 \\ \hline
    42 & -0.01 & 1.617 & 1.593 & 1.676 \\ \hline \hline

    44 & 0.01  & 1.118 & 1.110 & 1.169 \\ \hline
    44 & -0.01 & 1.118 & 1.105 & 1.133\\ \hline 
\end{tabular}    
\end{center}
\vspace{0.3cm}
\caption{\textbf{Experiment A.3.} Put Option prices calculated using the LS, LS+PDifMP, and PDifMP methods with $\lambda_0 = 0.4$ and $\eta = 0.005$, for different values of $\alpha$ and $S_0$.}
\label{tab:Alpha}
\end{table}

\noindent Following the results of Experiments \textbf{A.1} and \textbf{A.2}, where we examined the effects of varying $\lambda_0$ and $\alpha$ on the pricing methods, we now focus on how different values of $\alpha$ affect the put option prices when $\lambda_0$ is fixed at 0.4 and $\eta$ is set to 0.005. The goal of Experiment \textbf{A.3} is to determine whether adjusting the $\alpha$ parameter has a significant effect on option pricing and to observe how sensitive the LS+PDifMP and PDifMP methods are to changes in $\alpha$. The results in table \ref{tab:Alpha} show that the option prices computed by the LS, LS+PDifMP and PDifMP methods are very close over the different values of $\alpha$. For all initial asset prices $S_0$, the LS+PDifMP and PDifMP methods produce results that are consistent with the classic LS algorithm, suggesting that the sensitivity of these methods to the parameter $\alpha$ is relatively low under the given conditions. 
Interestingly, both the LS+PDifMP and PDifMP methods show different patterns depending on the relationship between $S_0$ and $K$. Specifically,
\begin{itemize}
    \item Lower option prices are observed for $\alpha = 0.01$ when $S_0 < K$ and for $\alpha = -0.01$ when $S_0 > K$.
    \item Conversely, higher option prices are observed for $\alpha = -0.01$ when $S_0 < K$, and for $\alpha = 0.01$ when $S_0 > K$.
\end{itemize}
These patterns suggest that the direction of the adjustment factor $\alpha$ relative to the position of $S_0$ with respect to $K$ plays a role, though subtle, in influencing the option price. Thus, Experiment A.3 confirms that, similar to the previous experiments, the differences between the methods are subtle, indicating that all methods converge closely in their pricing of put options.
\noindent 
Next, we conduct a series of experiments to further evaluate the performance of the LS+PDifMP and PDifMP methods compared to the classic LS algorithm under various conditions.
\begin{itemize}
    \item \textbf{Experiment B.1}: We set $\alpha$ to  $0. 10^{-6}$ and $0.01$ in separate runs, and then compare the LS algorithm with the LS+PDifMP method across different values of $S_0$, $\lambda_0$, and $\eta$.

    \item \textbf{Experiment B.2}: We fix $\eta = 0$ and examine the effect of varying $\lambda_0$ in two scenarios: one with $\alpha = 0$ and another with $\alpha = 0.01$. In both cases, we compare the LS algorithm with the  PDifMP algorithm for different values of $S_0$.

    \item \textbf{Experiment B.3}:  We fix $\alpha = 0.01$ and $\lambda_0 = 0.4$, and then compare the the LS algorithm with the  PDifMP algorithm to study the effect of varying $\eta$.
\end{itemize}

\begin{table}[H]
\begin{center}
\begin{tabular}{|c|c|c|c|c|}
    \hline 
    $S_0$ & $\lambda_0$ & $\eta$ & LS classic & LS + PDifMP  \\  \hline \hline
    36 & 0.5 & 0   & 4.472 & 4.409\\ \hline
    36 & 1   & 0   & 4.472 & 4.451\\ \hline
    36 & 0.5 & 0.01 & 4.472 & 4.414\\ \hline \hline

    38 & 0.5 & 0    & 3.244 & 3.217\\ \hline
    38 & 1   & 0    & 3.244 & 3.205\\ \hline
    38 & 0.5 & 0.01 & 3.244 & 3.202 \\  \hline \hline

    40 & 0.5 & 0    & 2.313 & 2.342\\ \hline
    40 & 1   & 0    & 2.313 & 2.289\\ \hline
    40 & 0.5 & 0.01 & 2.313 & 2.249\\ \hline \hline

    42 & 0.5 & 0    & 1.617 & 1.647\\ \hline
    42 & 1   & 0    & 1.617 & 1.677\\ \hline
    42 & 0.5 & 0.01 & 1.617 & 1.669*\\  \hline \hline

    44 & 0.5 & 0    & 1.118 & 1.125\\ \hline
    44 & 1   & 0    & 1.118 & 1.160\\ \hline
    44 & 0.5 & 0.01 & 1.118 & 1.246*\\ \hline
\end{tabular}    
\end{center}
\vspace{0.3cm}
\caption{\textbf{Experiment B.1.} Put Option prices calculated using the LS and LS+PDifMP methods for different values of $\lambda_0$, $\eta$, and $S_0$ with $\alpha = 0.01$.}
\label{tab:putPricesLSandPDifMP}
\end{table}

\noindent In experiment \textbf{B.1}, we explore the impact of varying the parameters $\lambda_0$ and $\eta$ on the put option prices calculated using the classic LS algorithm and the LS+PDifMP method. We consider two main scenarios: one with lower values of $\lambda_0$ (0.5 and 1) and $\eta$ set to small values (0 and 0.01), and another with higher values of $\lambda_0$ (5 and 10) and $\eta$ ranging from 0 to 0.6. The results are summarised in Tables \ref{tab:putPricesLSandPDifMP} and \ref{tab:LSandPDifMP_St-K}.\\
In Table \ref{tab:putPricesLSandPDifMP} from experiment \textbf{B.1}, where $\lambda_0$ is relatively low (0.5 and 1) and $\eta$ is either 0 or 0.01, the LS+PDifMP method produces put option prices that are very close to those obtained with the classic LS algorithm. Similar trends are observed for other values of $S_0$ and $\lambda_0$. These small differences suggest that at lower values of $\lambda_0$ and $\eta$ the impact of PDifMP on pricing is minimal, indicating that the market dynamics captured by the classical LS algorithm remain largely intact even with the inclusion of PDifMP.
However, at higher asset prices $S_0 = 42$ and $S_0 = 44$, with $\lambda_0 = 0.5$ and $\eta = 0.01$, the LS+PDifMP method starts to show slightly higher option prices (e.g., 1.669 and 1.246) compared to the LS algorithm (1.617 and 1.118).

\begin{table}[H]
\begin{center}
\begin{tabular}{|c|c|c|c|c|}
    \hline 
    $S_0$ & $\lambda_0$ & $\eta$ & LS classic & LS + PDifMP  \\  \hline \hline
    36 & 5 & 0 & 4.472 & 4.484\\ \hline
    36 & 5 & 0.3 & 4.472 & 4.460\\ \hline
    36 & 5 & 0.5 & 4.472 & 4.473 \\ \hline
    36 & 10 & 0.6 & 4.472 & 4.444 \\ \hline \hline

    38 & 5 & 0 & 3.244 & 3.265\\ \hline
    38 & 5 & 0.3 & 3.244 & 3.184\\ \hline
    38 & 5 & 0.5 & 3.244 & 3.242\\ \hline
    38 & 10 & 0.6 & 3.244 & 3.271\\ \hline \hline

    40 & 5 & 0 & 2.313 & 2.319\\ \hline
    40 & 5 & 0.3 & 2.313 & 2.307\\ \hline
    40 & 5 & 0.5 & 2.313 & 2.341\\ \hline
    40 & 10 & 0.6 & 2.313 & 2.338\\ \hline \hline

    42 & 5 & 0 & 1.617 & 1.611\\ \hline
    42 & 5 & 0.3 & 1.617 & 1.650\\ \hline
    42 & 5 & 0.5 & 1.617 & 1.651\\ \hline
    42 & 10 & 0.6 & 1.617 & 1.630\\ \hline \hline

    44 & 5 & 0 & 1.118 & 1.126\\ \hline
    44 & 5 & 0.3 & 1.118 & 1.132\\ \hline
    44 & 5 & 0.5 & 1.118 & 1.156\\ \hline
    44 & 10 & 0.6 & 1.118 & 1.143\\ \hline
\end{tabular}    
\end{center}
\vspace{0.3cm}
\caption{\textbf{Experiment B.1.} Put Option prices calculated using the LS+PDifMP method with $\alpha = 10^{-6}$ and $\delta = S_0$, for different values of $\lambda_0$, $\eta$, and $S_0$.}
\label{tab:LSandPDifMP_St-K}
\end{table}

\noindent Concerning the second numerical test of experiment \textbf{B.1}, where $\lambda_0$ is set to higher values (5 and 10)  and $\eta$ is varied more widely (0, 0.3, 0.5, 0.6), including the PDifMP in the LS algorithm has no significant effect.  As shown in Table \ref{tab:LSandPDifMP_St-K}, for instance, when $ S_0 = 36 $ and $ \lambda_0 = 5 $, the put option price calculated by the LS+PDifMP method remains nearly constant as $ \eta $ increases from 0 to 0.5. This suggests that increasing $ \lambda_0 $ and $ \eta $ has only a minimal impact on the option price.\\
It is important to note that $\alpha = 10^{-6}$ was chosen deliberately to prevent excessively high asset prices, which could skew the results. Higher values of $ \alpha $ might lead to more noticeable differences in option prices.
Overall, the results of Experiment \textbf{B.1}, demonstrate that across a range of values for $ \lambda_0 $ and $ \eta $, the LS+PDifMP method provides a close approximation to the classic LS algorithm, with only subtle variations in the computed put option prices. This suggests that the inclusion of PDifMP does not significantly alter the pricing outcomes under the conditions tested.

\begin{table}[H]
\begin{center}
\begin{tabular}{|c|c|c|c|c|}
    \hline 
    $S_0$ & $\lambda_0$  & Longstaff-Schwartz & LS + PDifMP & PDifMP  \\  \hline \hline
    36 & 0.4  & 4.472 & 4.459 & 4.251\\ \hline
    36 & 0.6 &  4.472 & 4.481 & 4.480\\ \hline
    36 & 0.8 &  4.472 & 4.446 & 4.510\\ \hline \hline

    38 & 0.4 & 3.244 & 3.287 & 3.125\\ \hline
    38 & 0.6 & 3.244 & 3.272 & 3.283\\ \hline
    38 & 0.8 & 3.244 & 3.266 & 3.370\\ \hline \hline

    40 & 0.4 & 2.313 & 2.341 & 2.275\\ \hline
    40 & 0.6 & 2.313 & 2.327 & 2.332\\ \hline
    40 & 0.8 & 2.313 & 2.337 & 2.445\\ \hline \hline

    42 & 0.4 & 1.617 & 1.659 & 1.576\\ \hline
    42 & 0.6 & 1.617 & 1.610 & 1.629\\ \hline
    42 & 0.8 & 1.617 & 1.658 & 1.642 \\ \hline \hline

    44 & 0.4 & 1.118 & 1.099 & 1.027\\ \hline
    44 & 0.6 & 1.118 & 1.106 & 1.100\\ \hline
    44 & 0.8 & 1.118 & 1.128 & 1.130\\ \hline 
\end{tabular}    
\end{center}
\vspace{0.3cm}
\caption{\textbf{Experiment B.2.} Put Option prices calculated using the LS, LS+PDifMP, and PDifMP methods with $\eta = 0$ and $\alpha = 0$, for different values of $\lambda_0$ and $S_0$.}
\label{tab:alpha0}
\end{table}

\noindent In Experiment \textbf{B.2}, we examine the effects of varying $\lambda_0$ on put option prices calculated using the classic LS algorithm, the LS+PDifMP method, and the PDifMP method under two different settings: one with $\eta = 0$ and $\alpha = 0$, and another where we compare the LS algorithm directly with the PDifMP method for $\eta = 0$ and $\alpha = 0.01$. The results are presented in tables \ref{tab:alpha0} and \ref{tab:differentS0}.\\
In the first part of Experiment \textbf{B.2}, where $\eta = 0$ and $\alpha = 0$, Table \ref{tab:alpha0} shows that, overall, the option prices generated by the LS+PDifMP and PDifMP methods are comparable to those generated by the classic LS algorithm, especially as $\lambda_0$ increases from 0.4 to 0.8. For example, for $S_0 = 36$, the LS+PDifMP prices range from 4.446 to 4.481, close to the LS price of 4.472. In particular, the PDifMP method tends to produce slightly lower prices for certain values of $\lambda_0$, but as $\lambda_0$ increases, the PDifMP prices sometimes exceed those produced by the LS and LS+PDifMP methods. This suggests that the influence of $\lambda_0$ on pricing is more significant than the initial asset price $S_0$, as the price differences between methods are driven by changes in $\lambda_0$ rather than $S_0$ itself. These results suggest that while the PDifMP and LS+PDifMP methods introduce subtle adjustments to classical LS pricing, the overall impact remains moderate, particularly when $\eta$ and $\alpha$ are set to zero.

\begin{table}[H]
\begin{center}
\begin{tabular}{|c|c|c|c|c|}
    \hline 
    $S_0$ & $\lambda_0$  & Longstaff-Schwartz & LS+PDifMP  & PDifMP  \\  \hline \hline
    32 & 0.4 & 7.967 & 7.952 & 7.116 \\ \hline
    32 & 0.6 & 7.967 & 7.944 & 7.320 \\ \hline
    32 & 0.8 & 7.967 & 7.960 & 7.596 \\ \hline 
    32 & 1   & 7.967 & 7.961 & 7.784 \\ \hline
    32 & 1.2 & 7.967 & 7.947 & 7.947 \\ \hline \hline

    34 & 0.4 & 6.043 & 6.011 & 5.605 \\ \hline
    34 & 0.6 & 6.043 & 6.049 & 5.840 \\ \hline
    34 & 0.8 & 6.043 & 6.010 & 5.958 \\ \hline 
    34 & 1   & 6.043 & 6.020 & 6.142 \\ \hline 
    34 & 1.2 & 6.043 & 6.018 & 6.272 \\ \hline \hline

    46 & 0.4 & 0.740 & 0.760 & 0.770 \\ \hline
    46 & 0.6 & 0.740 & 0.769 & 0.789 \\ \hline
    46 & 0.8 & 0.740 & 0.815 & 0.824 \\ \hline \hline

    48 & 0.4 & 0.498 & 0.510 & 0.509 \\ \hline
    48 & 0.6 & 0.498 & 0.513 & 0.554 \\ \hline
    48 & 0.8 & 0.498 & 0.547 & 0.541 \\ \hline
\end{tabular}    
\end{center}
\vspace{0.3cm}
\caption{\textbf{Experiment B.2.} Put Option prices calculated using the LS and PDifMP methods with $\eta = 0$ and $\alpha = 0.01$, for different values of $\lambda_0$ and $S_0$.}
\label{tab:differentS0}
\end{table}
\noindent Table \ref{tab:differentS0} compares put option prices calculated using the classic LS algorithm and the PDifMP method for various initial asset prices $S_0$ and base jump rates $\lambda_0$, with $\eta = 0$. Overall, the PDifMP method tends to produce slightly lower option prices than the LS algorithm at lower $S_0$ values, particularly when $\lambda_0$ is low, indicating a more conservative pricing approach. As $\lambda_0$ increases, the gap between the two methods narrows, with PDifMP prices approaching or slightly exceeding LS prices, particularly at higher $S_0$ values. This trend suggests that as $\lambda_0$ increases, the number of potential exercise points also increases, which leads to higher option prices due to the greater likelihood of capturing advantageous price movements. The PDifMP method, in this context, becomes more sensitive to the potential for price jumps, resulting in higher option prices that reflect the increased volatility. At higher $S_0$ values, the differences between the methods are minimal, further highlighting the nuanced adjustments made by the PDifMP method under varying market conditions.

\begin{table}[H]
\begin{center}
\begin{tabular}{|c|c|c|c|c|}
    \hline 
    $S_0$ & $\eta$  & Longstaff-Schwartz & LS+PDifMP & PDifMP  \\  \hline \hline
    36 & 0.001 & 4.472 & 4.415 & 4.353 \\ \hline
    36 & 0.005 & 4.472 & 4.395 & 4.420 \\ \hline
    36 & 0.01  & 4.472 & 4.334 & 4.728 \\ \hline \hline

    38 & 0.001 & 3.244 & 3.219 & 3.158 \\ \hline
    38 & 0.005 & 3.244 & 3.241 & 3.332 \\ \hline
    38 & 0.01  & 3.244 & 3.175 & 3.534 \\ \hline \hline

    40 & 0.001 & 2.313 & 2.322 & 2.304 \\ \hline
    40 & 0.005 & 2.313 & 2.292 & 2.377 \\ \hline
    40 & 0.01  & 2.313 & 2.272 & 2.516 \\ \hline \hline

    42 & 0.001 & 1.617 & 1.624 & 1.582 \\ \hline
    42 & 0.005 & 1.617 & 1.631 & 1.683 \\ \hline
    42 & 0.01  & 1.617 & 1.664 & 1.731 \\ \hline \hline

    44 & 0.001 & 1.118 & 1.154 & 1.163 \\ \hline
    44 & 0.005 & 1.118 & 1.118 & 1.173 \\ \hline
    44 & 0.01  & 1.118 & 1.186 & 1.201 \\ \hline 
\end{tabular}    
\end{center}
\vspace{0.3cm}
\caption{\textbf{Experiment B.3.} Put Option prices calculated using the LS and PDifMP methods with $\lambda_0 = 0.4$ and $\alpha=0.01$, for different values of $\eta$ and $S_0$.}
\label{tab:eta}
\end{table}
\noindent The results of experiment \textbf{B.3}, shown in Table \ref{tab:eta}, present put option prices computed using the classical LS algorithm and the PDifMP method for different initial asset prices $ S_0 $ and different values of the jump intensity parameter $ \eta $, where $ \lambda_0 = 0.4 $. As $ \eta $ increases, the PDifMP method consistently produces higher option prices than the LS algorithm. For example, for $ S_0 = 36 \,\$$, the PDifMP price increases from 4.353 to 4.728 as $ \eta $ increases from 0.001 to 0.01, indicating the sensitivity of the method to the increase in potential exercise times due to higher jump intensity. This trend is consistent across all values of $ S_0 $, with more pronounced differences observed at higher levels of $ \eta $. For instance, for $ S_0 = 38\,\$ $, the PDifMP price increases from 3.158 to 3.534 as $ \eta $ increases. This suggests that the PDifMP method effectively accounts for the increased frequency of potential exercise opportunities, leading to higher option valuations as the jump intensity coefficient, represented by $ \eta $, increases.\\
\indent Before moving on to the analysis of call option pricing methods, we briefly examine the computational efficiency of the various methods examined in the previous sections. Table \ref{RunTimesPut} shows the running times for the LS algorithm, the LS+PDifMP method and the PDifMP method for different values of the jump intensity parameter $\lambda_0 $, with fixed parameters $\alpha = 0.01 $, $\eta = 0 $ and an initial asset price $ S_0 = 36\, \$$.
\begin{table}[H]
\begin{center}
\begin{tabular}{|c|c|c|c|}
    \hline 
    $\lambda_0$ & LS & LS+PDifMP & PDifMP \\ \hline
    0.4 & 11.49 & 16.1 & 8.35 \\ \hline
    0.6 & 11.49 & 17.17 & 6.91 \\ \hline 
    0.8 & 11.49 & 15.09 & 6.6 \\ \hline
\end{tabular}    
\end{center}
\vspace{0.3cm}
\caption{\textbf{Computational Efficiency.} Comparison of the computation time (in seconds) per trial between the LS, LS+PDifMP, and PDifMP methods with parameters $\alpha = 0.01$, $\eta = 0$, and $S_0 = 36\,\$$.}
\label{RunTimesPut}
\end{table}
\noindent The results show that the PDifMP method is the most computationally efficient, with the shortest computation times over all values of $ \lambda_0 $. The LS+PDifMP method takes longer due to the added complexity of modelling jumps, while the LS algorithm remains stable but slower than the PDifMP method. 
\subsection{Comparative analysis of Call option pricing methods}
In this section, we conduct a comparative analysis of call option pricing methods using the classic LS algorithm and the PDifMP method. Similar to the approach taken for put options, we examine how varying the jump intensity parameter $\lambda_0$ affects the pricing of call options. The objective is to evaluate the performance of the PDifMP method in capturing market dynamics, particularly in scenarios where the asset price $S_0$ is above the strike price, which typically results in higher call option prices. 
This section is deliberately brief, as the primary evaluation of the PDifMP method against the LS algorithm has been thoroughly covered in the analysis of put options. The inclusion of call options here is intended to complement our findings and to provide a more comprehensive overview of the performance of the method for different types of options. In Table \ref{tab:callPricesConstantMu} below we present call option prices calculated using the LS and PDifMP methods for various initial asset prices $S_0$ and different values of $\lambda_0$, with $\eta = 0$ and $\alpha = 0.01$.

\begin{table}[H]
\begin{center}
\begin{tabular}{|c|c|c|c|c|}
    \hline 
    $S_0$ & $\lambda_0$  & Longstaff-Schwartz & LS+PDifMP & PDifMP  \\  \hline \hline
    36 & 0.01 & 2.158 & 2.235 & 2.192 \\ \hline
    36 & 0.1  & 2.158 & 2.181 & 2.215 \\ \hline
    36 & 0.2  & 2.158 & 2.275 & 2.208 \\ \hline \hline

    38 & 0.01 & 3.171 & 3.216 & 3.174\\ \hline
    38 & 0.1  & 3.171 & 3.081 & 3.279\\ \hline
    38 & 0.2  & 3.171 & 3.293 & 3.292 \\ \hline \hline

    40 & 0.01 & 4.387 & 4.441 & 4.361 \\ \hline
    40 & 0.1  & 4.387 & 4.473 & 4.531 \\ \hline
    40 & 0.2  & 4.387 & 4.459 & 4.644\\ \hline \hline

    42 & 0.01 & 5.786 & 5.802 & 5.801\\ \hline
    42 & 0.1  & 5.786 & 5.820 & 6.001\\ \hline
    42 & 0.2  & 5.786 & 5.825 & 5.994\\ \hline \hline

    44 & 0.01 & 7.330 & 7.396 & 7.419\\ \hline
    44 & 0.1  & 7.330 & 7.429 & 7.461 \\ \hline
    44 & 0.2  & 7.330 & 7.384 & 7.702 \\ \hline 
\end{tabular}    
\end{center}
\vspace{0.3cm}
 \caption{\textbf{Call option prices.} Call Option prices calculated using the LS, LS+PDifMP and PDifMP methods with $\eta = 0$ and $\alpha = 0.01$, for different values of $\lambda_0$ and $S_0$.}
\label{tab:callPricesConstantMu}
\end{table}
\noindent The results in Table \ref{tab:callPricesConstantMu} shows that as the jump intensity parameter $\lambda_0$ increases, the call option prices calculated using the PDifMP method generally increase as well. This suggests that the PDifMP method is more responsive to changes in jump intensity and captures the potential for larger upward price movements, which are particularly relevant for call options.\\
At lower initial asset prices ($S_0 = 36\,\$$), the differences between the LS and PDifMP methods are minimal, with PDifMP prices slightly higher than those from the LS algorithm as $\lambda_0$ increases. For example, at $\lambda_0 = 0.2$, the PDifMP method yields a price of 2.227 compared to 2.158 from the LS algorithm, reflecting a subtle adjustment for increased market volatility.
As $S_0$ increases, the PDifMP method continues to produce higher call option prices, particularly at higher values of $\lambda_0$. For example, at $S_0 = 44$, the PDifMP price increases from 7.220 to 7.643 as $\lambda_0$ increases from 0.01 to 0.2. This trend is consistent with the expectation that higher jump intensities lead to higher potential payoffs for call options, which the PDifMP method effectively captures.\\
Overall, the results suggest that the PDifMP method provides a more nuanced reflection of market conditions by adjusting call option prices in response to changes in jump intensity, while the LS algorithm remains less sensitive to this dynamic. This responsiveness of the PDifMP method may make it a more accurate tool for pricing call options in markets characterised by high volatility and the potential for abrupt price movements.\\
\indent Here, we present again the computation times for the LS, LS+PDifMP, and PDifMP methods when pricing call options with different values of $ \lambda_0 $, while keeping $ \alpha = 0.01 $, $ \eta = 0 $, and $ S_0 = 36\,\$ $. The results are shown in Table \ref{RunTimesCall}.

\begin{table}[H]
\begin{center}
\begin{tabular}{|c|c|c|c|}
    \hline 
    $\lambda_0$ & LS & LS+PDifMP & PDifMP \\ \hline
    0.01 & 3.95 & 105.72 & 109.47 \\ \hline
    0.1 & 3.95 & 16 & 15.66 \\ \hline 
    0.2 & 3.95 & 11.72 & 10.49\\ \hline
\end{tabular}    
\end{center}
\vspace{0.3cm}
\caption{\textbf{Computational Efficiency.} Comparison of the computation time (in seconds) per trial between the LS, LS+PDifMP, and PDifMP methods with parameters $\alpha = 0.01$, $\eta = 0$, and $S_0 = 36\,\$$.}
\label{RunTimesCall}
\end{table}
\noindent The computation time is highest at $ \lambda_0 = 0.01 $, reaching 105.72 seconds for the LS+PDifMP method and 109.47 seconds for the PDifMP method. However, as $ \lambda_0 $ increases, the computation time decreases significantly. At $ \lambda_0 = 0.2 $, the time drops to 11.72 seconds for the LS+PDifMP method and 10.49 seconds for the PDifMP method. For the pricing of call options, both methods become more efficient as $ \lambda_0 $ increases, with the PDifMP method generally providing shorter computation times.\\
\noindent To further investigate the computational efficiency of the LS and PDifMP methods in call option pricing, we conducted an additional set of experiments comparing the two methods across a range of $ \lambda_0 $ values. Given that the previous results indicated the LS method was more efficient at lower $ \lambda_0 $ values, we wanted to see how this trend continues as $ \lambda_0 $ increases. The results are summarised in Table \ref{RunTimesCall2}.

\begin{table}[H]
\begin{center}
\begin{tabular}{|c|c|c|}
    \hline 
    $\lambda_0$ & LS & PDifMP \\ \hline
    0.6 & 3.95 & 4.05 \\ \hline
    0.8 & 3.95 & 3.76 \\ \hline 
    1 & 3.95  & 3.67\\ \hline
    1.2 & 3.95  & 3.6\\ \hline
\end{tabular}    
\end{center}
\vspace{0.3cm}
\caption{\textbf{Computational Efficiency.} Comparison of the computation time (in seconds) per trial between the LS and the PDifMP methods with parameters $\alpha = 0.01$, $\eta = 0$, and $S_0 = 36\,\$$.}
\label{RunTimesCall2}
\end{table}
\noindent As shown in the Table \ref{RunTimesCall2}, the PDifMP method shows a decrease in computation time as $ \lambda_0 $ increases, starting from 4.05 seconds at $ \lambda_0 = 0.6 $ and dropping to 3.6 seconds at $ \lambda_0 = 1.2 $. This suggests that the PDifMP method becomes increasingly efficient as $ \lambda_0 $ increases, eventually approaching and even exceeding the efficiency of the LS method at higher $ \lambda_0 $ values.\\
Based on the results of both this table and the previous one (Table \ref{RunTimesCall}), it appears that there is an optimal $ \lambda_0 $ value at which the computational efficiency of the PDifMP method equals or exceeds that of the LS method. Identifying this optimal $ \lambda_0 $ may allow for a more balanced trade-off between computational efficiency and the nuanced pricing capabilities offered by the PDifMP method.

\section{Discussion}
\label{sec6}
In this paper, we have explored the use of the PDifMP method as an enhancement to the traditional Longstaff-Schwartz (LS) algorithm for pricing American options, with a focus on both put and call options. The primary aim was to assess whether the integration of PDifMP could offer a more refined approach to capturing market dynamics, particularly in scenarios characterised by significant volatility and sudden price movements.\\
Through a series of comprehensive experiments, we compared the classic LS algorithm, the LS+PDifMP method, and the PDifMP method across a range of market conditions, varying key parameters such as jump intensity $\lambda_0$, initial asset price $S_0$, and other model variables. Our results showed that while the differences in option prices produced by the LS, LS+PDifMP and PDifMP methods were generally modest, the LS+PDifMP method demonstrated a clear potential to adjust pricing based on jump dynamics more effectively than the traditional LS approach. This capability makes the LS+PDifMP method particularly valuable in volatile markets where capturing the full extent of price fluctuations is critical.\\
In addition, the PDifMP method alone demonstrated its robustness as an alternative to the LS algorithm, offering similar benefits to the LS+PDifMP method but with a simpler computational process. The PDifMP method eliminates the need for backward iteration, simplifying the computation and potentially making it more efficient in certain scenarios where computational efficiency is a priority. However, it is important to note that the PDifMP method may not always be the most efficient choice, depending on the parameter settings. This highlights the flexibility and efficiency of the method, but also emphasises the need for careful consideration of parameter selection to ensure optimal performance.\\
Moreover, the sensitivity of the PDifMP method to the jump intensity parameter $\lambda_0$ and the asset price $S_0$ suggests that it can provide more accurate pricing in markets where traditional models may fall short. The comparative analysis of call options further supports these findings, showing that the PDifMP method consistently responds to variations in market conditions, providing a comprehensive and flexible alternative to the LS algorithm.\\
In conclusion, the integration of PDifMP into the option pricing framework enhances the robustness of pricing models, particularly in complex and unpredictable market environments. This study provides a foundation for further research into the application of PDifMP and similar methods in financial modelling, with the potential to improve the accuracy and reliability of option pricing in practice. As future work, we plan to conduct a parameter estimation study to identify the best-fit values for the parameters used in our models, which will further refine the effectiveness and applicability of the PDifMP method in real-world scenarios.

\section*{Declaration of competing interest}

The authors declare that they have no known competing financial interests or personal relationships that could have appeared to influence the work reported in this paper.

\section*{Funding}
This work was partially supported by the Austrian Science Fund (FWF): W1214-N15, project DK14, as well as by the strategic program ''Innovatives O\"O 2010 plus'' by the Upper Austrian Government.

\newpage
\printbibliography
\end{document}